\documentclass[iop,apj,a4paper]{emulateapj}
\usepackage{apjfonts,amsmath,amssymb,epsf}
\submitted{Received 2012 August 2; accepted 2012 November 14; published 2012 December 26}
\journalinfo{THE ASTROPHYSICAL JOURNAL, 763:3 (12pp), 2013 January 20}

\shortauthors{
Kobayashi,
Inoue,
\& Inoue
}

\shorttitle{Revisiting the Cosmic Star Formation History}

\begin{document}

\title{REVISITING THE COSMIC STAR FORMATION HISTORY: CAUTION ON THE
UNCERTAINTIES IN DUST CORRECTION AND STAR FORMATION RATE CONVERSION}

%
%


\author{Masakazu A. R. Kobayashi$^{1,\ 2}$
Yoshiyuki Inoue$^3$, and
Akio K. Inoue$^4$}
\address{$^1$ Astronomy Data Center, National Astronomical
Observatory of Japan, Mitaka, Tokyo 181-8588, Japan; kobayashi@cosmos.phys.sci.ehime-u.ac.jp}
\address{$^2$ Research Center for Space and Cosmic Evolution, Ehime
University, Bunkyo-cho, Matsuyama 790-8577, Japan}
\address{$^3$ Kavli Institute for Particle Astrophysics and Cosmology,
Department of Physics and SLAC National Accelerator Laboratory,\\
Stanford University, Stanford, CA 94305, USA}
\address{$^4$ College of General Education, Osaka Sangyo University,
3-1-1 Nakagaito Daito, Osaka 574-8530, Japan}


\begin{abstract}

 The cosmic star formation rate density (CSFRD) has been
 observationally investigated out to redshift $z\simeq 10$.  However,
 most of the theoretical models for galaxy formation underpredict the
 CSFRD at $z\gtrsim 1$.  Since the theoretical models reproduce the
 observed luminosity functions (LFs), luminosity densities (LDs), and
 stellar mass density at each redshift, this inconsistency does not
 simply imply that theoretical models should incorporate some missing
 unknown physical processes in galaxy formation.  Here, we examine the
 cause of this inconsistency at UV wavelengths by using a mock catalog
 of galaxies generated by a semi-analytic model of galaxy formation.
 We find that this inconsistency is due to two observational
 uncertainties: the dust obscuration correction and the conversion
 from UV luminosity to star formation rate (SFR).  The methods for
 correction of obscuration and SFR conversion used in observational
 studies result in the overestimation of the CSFRD by $\sim
 0.1$--$0.3$~dex and $\sim 0.1$--$0.2$~dex, respectively, compared to
 the results obtained directly from our mock catalog.  We present new
 empirical calibrations for dust attenuation and conversion from
 observed UV LFs and LDs into the CSFRD.

\end{abstract}


 \section{Introduction}\label{section:intro}

 The cosmic star formation rate density (CSFRD), $\dot{\rho}_\star$,
 is one of the most fundamental quantities that reveals how present
 galaxies were formed and evolved in the universe.  It has been probed
 observationally since the seminal works of Lilly et al. (1996) and
 Madau et al. (1996, 1998) using various star formation rate (SFR)
 indicators such as luminosities of the stellar continuum at the
 rest-frame ultraviolet (UV) and nebular emission lines (e.g.,
 H$\alpha$).  Various observational results are compiled in Hopkins
 (2004; H04) and Hopkins \& Beacom (2006; HB06), in which the
 cosmology, stellar initial mass function (IMF), and dust obscuration
 correction are unified.  The best-fit CSFRD function using these
 results has been widely utilized not only in observational studies
 (e.g., Karim et al. 2011) but also in theoretical studies (e.g.,
 Coward et al. 2008; Kistler et al. 2009; Tominaga et al. 2011; Wang
 \& Dai 2011).

 The CSFRD at $z\lesssim 1$ has been confirmed by various SFR
 indicators obtained from wide-field surveys such as the
 \textit{Galaxy Evolution Explorer} (e.g., Wyder et al. 2005; Robotham
 \& Driver 2011).  Its characteristic feature is a rapid increase with
 redshift ($\dot{\rho}_\star \propto (1+z)^{3-4}$ at $z\lesssim 1$;
 H04 and references therein).  Thus, the CSFRD at $z\sim 1$ is an
 order of magnitude higher than that in the local universe.  In
 contrast, the CSFRD is still uncertain in the higher redshift range
 (i.e., $z\gtrsim 1$) where popular SFR indicators include the
 rest-frame UV continuum stellar emission and infrared (IR) dust
 emission.  This uncertainty is due to uncertainties in the estimation
 of the CSFRD from observed data: the dust obscuration correction for
 the UV continuum, contamination from the old stellar population to
 the IR luminosity, estimation of the total IR luminosity, the
 faint-end slope of the luminosity function (LF), and the conversion
 factor from luminosity into SFR.  These uncertainties result in the
 well-known inconsistencies in some physical quantities between direct
 measurements and values inferred from the HB06 CSFRD function such as
 stellar mass density (SMD; $\rho_\star$: e.g., Wilkins et al. 2008;
 Choi \& Nagamine 2012; Benson 2012), core-collapse supernova rate
 (e.g., Horiuchi et al. 2011; see also Botticella et al. 2012 who
 report there is no inconsistency in the local 11 Mpc volume), and
 extragalactic background light (e.g., Raue \& Meyer 2012).

 The CSFRD has also been calculated theoretically using galaxy
 formation models: hydrodynamic simulation (e.g., Nagamine et
 al. 2006) or semi-analytic models (e.g., Cole et al. 2001; Nagashima
 \& Yoshii 2004; Benson 2012).  In these models, a galaxy-by-galaxy
 basis calculation is executed based on a detailed hierarchical
 structure formation scenario.  Therefore, the CSFRD at a certain
 redshift can be calculated by simply integrating the SFR of each
 galaxy at that redshift.  Although these theoretical models reproduce
 reasonably well the observed LFs and the luminosity densities (LDs)
 at the rest-frame wavelength dominated by stellar emission, and the
 SMD at both local and high $z$, most underpredict the CSFRD compared
 to that estimated observationally (e.g., Nagashima \& Yoshii 2004;
 Nagashima et al. 2005; Lacey et al. 2011; Benson 2012).  While the
 underprediction of the CSFRD might be attributed to some missing
 unknown physical processes of galaxy formation and evolution, it is
 worth investigating whether or not the uncertainties in estimating
 the CSFRD from directly observed data could be the origin of the
 disagreement.

 \begin{figure*}
  \epsscale{0.73}
  \plotone{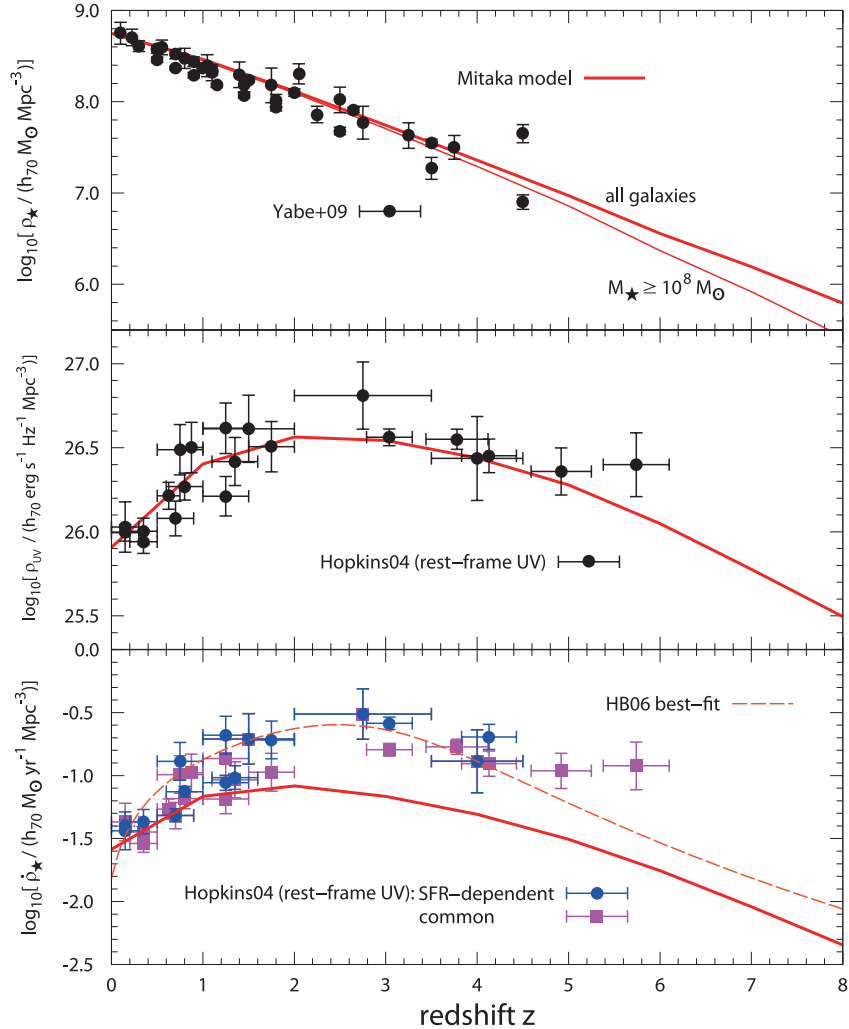}
  \caption{
  Redshift evolution of the SMD (top), the rest-frame UV (i.e.,
  $\lambda = 1500$--2800~{\AA}) LD (middle), and the CSFRD (bottom).
  The solid curves in each panel are the predictions of the Mitaka
  model.  The thin solid curve in the top panel is the SMD for model
  galaxies with $M_\star \ge 10^8\ M_\odot$.  The dashed curve in the
  bottom panel is the best-fit Cole et al. (2001) functional form to
  the H04 $\dot{\rho}_\star$.  The symbols with error bars are
  observational data.  The observational data shown in the top panel
  are compiled in Yabe et al. (2009) and those in the middle and
  bottom panels are from H04, respectively.  In the bottom panel, the
  boxes (circles) are evaluated using a common (SFR-dependent)
  obscuration correction from H04 for the same observed data plotted
  in the middle panel.
  }
  \label{fig-LD+CSFR}
 \end{figure*}
 Here, we examine the CSFRD through a comparison of the observational
 data compiled in H04 with the mock catalog of galaxies generated by
 one of the semi-analytic models for galaxy formation, the so-called
 Mitaka model (Nagashima \& Yoshii 2004; see also Nagashima et
 al. 2005).  The Mitaka model reproduces various kinds of observations
 not only for local galaxies including the stellar continuum LFs
 (Nagashima \& Yoshii 2004) but also for high-$z$ Lyman-break galaxies
 and Ly$\alpha$ emitters (Kashikawa et al. 2006; Kobayashi et
 al. 2007, 2010).  In this paper, we focus on the rest-frame UV
 continuum luminosity as an SFR indicator; this is usually applied to
 galaxies in the high-$z$ universe (e.g., Madau et al. 1996, 1998).
 Investigation of other SFR indicators, such as the rest-frame IR
 continuum emitted by interstellar dust, will be the subject of future
 work.

 As shown in the top and middle panels of Figure~\ref{fig-LD+CSFR},
 the Mitaka model reproduces well the measured SMD, $\rho_\star$, and
 LD at rest-frame UV wavelengths, $\rho_\mathrm{UV}$, which have not
 been corrected for interstellar dust attenuation, in the redshift
 range $z = 0$--$6$.  However, the model prediction for the CSFRD is
 underestimated at $z\gtrsim 1$ by a factor of $\sim 2$--3 (i.e.,
 $\sim 0.3$--0.5~dex) relative to the median value of
 $\dot{\rho}_\star$ compiled in H04 as shown in the bottom panel of
 Figure~\ref{fig-LD+CSFR}.  It should be emphasized that these H04
 CSFRD data are calculated using the same observational UV LDs plotted
 in the middle panel of Figure~\ref{fig-LD+CSFR}, and which are
 reasonably reproduced by our model.  Moreover, our CSFRD is
 consistent with the upper limit for $\dot{\rho}_\star$ given by
 Strigari et al. (2005) estimated from an upper limit on the diffuse
 supernova neutrino background with Super-Kamiokande.  In this paper
 we show that the underestimation of the Mitaka model relative to the
 H04 CSFRD can be fully attributed to two observational uncertainties,
 the dust obscuration correction and the SFR conversion used in
 observational studies.  Therefore the underestimation using
 theoretical galaxy formation models relative to the H04 CSFRD does
 not necessarily imply that these models should incorporate some
 missing unknown physical processes.

 \begin{figure*}
  \epsscale{.98}
  \plotone{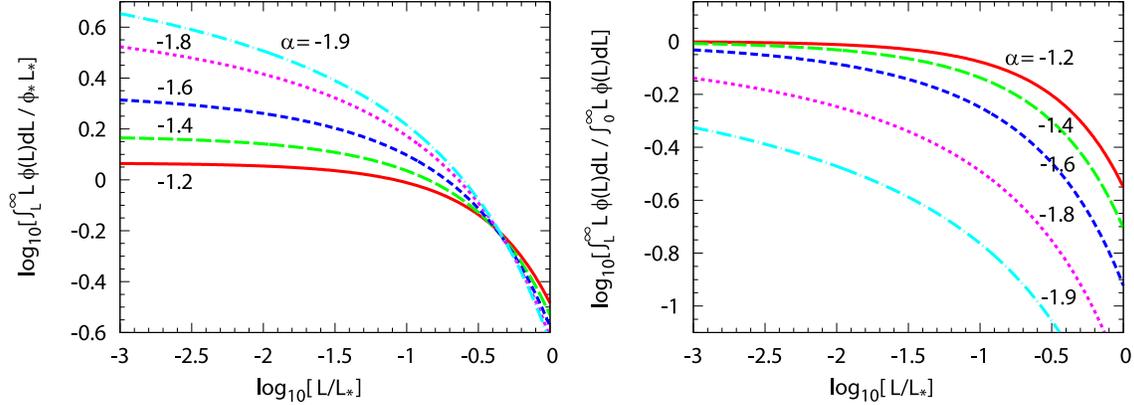}
  \caption{
  LD contribution from galaxies brighter than the horizontal axis of
  $L/L_\ast$ calculated using the Schechter function with various
  faint-end slopes of the LF.  As on the figure, the solid,
  long-dashed, short-dashed, dotted, and dash-dotted curves are for
  $\alpha = -1.2$, $-1.4$, $-1.6$, $-1.8$, and $-1.9$, respectively.
  The left and right panels are the LD normalized by $\phi_\ast
  L_\ast$ and the total LD, respectively.
  }
  \label{fig-Schechter}
 \end{figure*}
 The paper is organized as follows.  In Section~\ref{sec:ObsUnc}, we
 list the uncertainties in estimating the CSFRD from observed UV LFs.
 In Section~\ref{sec:Model}, we describe the key prescriptions of the
 Mitaka model relating to the observational uncertainties in
 estimating the CSFRD.  Then we compare the model results with the
 observed LFs, LDs, and CSFRD in the redshift range $z=0$--10 in
 Section~\ref{sec:Results}.  We summarize our work in
 Section~\ref{sec:Summary+Discussion}, where we also provide a
 discussion which includes a new formula for the obscuration
 correction at rest-frame UV wavelengths in
 Section~\ref{sec:Summary+Discussion}.  Throughout this paper, we
 adopt the 737 cosmology, i.e., $H_0=70~\mathrm{km\ s^{-1}\ Mpc^{-1}}$
 (i.e., $h_{70}\equiv h/0.7 = 1$), $\Omega_M=0.3$, and $\Omega_\Lambda
 =0.7$, and a Salpeter IMF with a mass range of $0.1$--$60~M_\odot$.
 All magnitudes are expressed in the AB system and the wavelengths are
 given in the rest frame unless otherwise stated.
 \section{Uncertainties in Estimating the Cosmic Star Formation Rate
 Density}\label{sec:ObsUnc}

 In the process of estimating the CSFRD at a certain redshift from
 observational data of galaxies, the basic direct observable
 quantities are the luminosities which can be used as SFR indicators.
 Luminosities can be converted into the CSFRD via the following
 processes.  First, the observed luminosities are corrected for dust
 attenuation in order to obtain intrinsic luminosities which are
 expected to correlate more directly with the SFR.  Next, the LF is
 constructed from the data corrected for dust obscuration.  Since the
 data are flux-limited samples, the main uncertainty in the shape of
 the LF lies in the faint-end slope index.  Next, the LD is derived by
 integrating the LF.  These processes can be interchanged if a
 luminosity-independent obscuration correction is adopted.  Finally,
 the LD is converted to the CSFRD using the SFR conversion factor.
 Thus, uncertainties may arise from (1) the faint-end slope of the LF,
 (2) the conversion factor from luminosity into SFR, and (3) the dust
 obscuration correction.

 There are other uncertainties in the estimation of the CSFRD such as
 the limiting luminosity to which LF is integrated (e.g., Reddy \&
 Steidel 2009), the IMF (e.g., HB06), and leakage of ionizing photons
 (e.g., Rela\~{n}o et al. 2012).  As the limiting luminosity and IMF
 are calibrated in a common fashion in H04, we do not need to examine
 these uncertainties here.  We can also neglect the uncertainty caused
 by the leakage of ionizing photons because we treat the UV continuum
 luminosity as an SFR indicator.  Hence, here we briefly describe the
 three uncertainties described in the previous paragraph, focusing on
 the UV continuum luminosity as an SFR indicator.

  \subsection{Faint-end Slope of the Luminosity Function}

  In general, the uncertainty in the faint-end slope of the LF,
  $\alpha$, leads that of the integrated LD.  It is more significant
  at high $z$ simply because only bright galaxies are detected at high
  $z$ in flux-limited surveys and thus the uncertainty in $\alpha$
  becomes larger.

  Figure~\ref{fig-Schechter} shows the LDs for various $\alpha$ as a
  function of minimum luminosity.  We normalize the LDs by the
  characteristic luminosity $L_\ast$ and number density $\phi_\ast$
  and adopt the Schechter function form for the LF.  For a typical
  value of $\alpha \sim -1.5$ in the UV LFs at $z=0$--6 (e.g., Oesch
  et al. 2010), the uncertainty of $\Delta \alpha = \pm 0.3$ results
  in $\sim 0.2$--0.3~dex (a factor of $\sim 1.6$--2) uncertainty in
  the LD integrated to $\log_{10}\left(L / L_\ast \right) \sim -3$.
  If this uncertainty fluctuates randomly to positive or negative
  values, it may be the origin of the dispersion in the LDs at each
  redshift as shown in the middle panel of Figure~\ref{fig-LD+CSFR}.

  However, because the UV LD calculated from the Mitaka model lies
  around the median value of the observed UV LDs, the uncertainty in
  $\alpha$ is not the origin of the systematic underestimation of the
  CSFRD calculated using the Mitaka model compared to that of H04;
  hence, we do not treat this uncertainty further here.

  \subsection{Conversion from Stellar Continuum Luminosity into SFR}

  The intrinsic (i.e., without dust attenuation) stellar continuum
  luminosity at UV wavelengths, $L_{\nu, \mathrm{UV}}^\mathrm{int}$,
  is one of the best SFR indicators.  This is because it is dominated
  by stellar radiation from recently formed massive stars and
  rest-frame UV photons from redshift $z=3$--10 to optical or near-IR
  in the observer frame at which we can observe most efficiently with
  current telescopes.

  In observational studies, the conversion factor from $L_{\nu,
  \mathrm{UV}}^\mathrm{int}$ to SFR, $C_\mathrm{SFR}$, given by
  Kennicutt (1998; hereafter, K98) has been widely utilized:
  \begin{displaymath}
   C_\mathrm{SFR}^\mathrm{K98} \equiv \mathrm{SFR}/ L_{\nu,
    \mathrm{UV}}^\mathrm{int} = 1.4\times 10^{-28}\ M_\odot\
    \mathrm{yr^{-1}\ (erg\ s^{-1}\ Hz^{-1})^{-1}}.
  \end{displaymath}
  H04 and HB06 also
  adopted $C_\mathrm{SFR}^\mathrm{K98}$.  However, it should be noted
  that, as explicitly stated in K98, the linear relation holds only
  for galaxies which continuously form stars over timescales of
  100~Myr or longer, have solar metallicity, and the Salpeter IMF in
  the mass range $0.1$--$100~M_\odot$ (see also Madau et
  al. 1998).\footnote{While this mass range for the IMF is different
  from that adopted in our model, the resultant difference in $L_{\nu,
  \mathrm{UV}}^\mathrm{int}$ (and hence in $C_\mathrm{SFR}$) at
  $\lambda = 1500$~{\AA} is found to be negligibly small (i.e.,
  $\lesssim 0.01$~dex) except during the very young phase of star
  formation (i.e., $\log_{10}(t / \mathrm{yr}) \lesssim 6.5$)
  according to the time evolution of $L_{\nu,
  \mathrm{UV}}^\mathrm{int}$ for the simple stellar population
  calculated with the \textit{PEGASE} population synthesis model (Fioc
  \& Rocca-Volmerange 1997).}

  \begin{figure}
   \epsscale{.95}
   \plotone{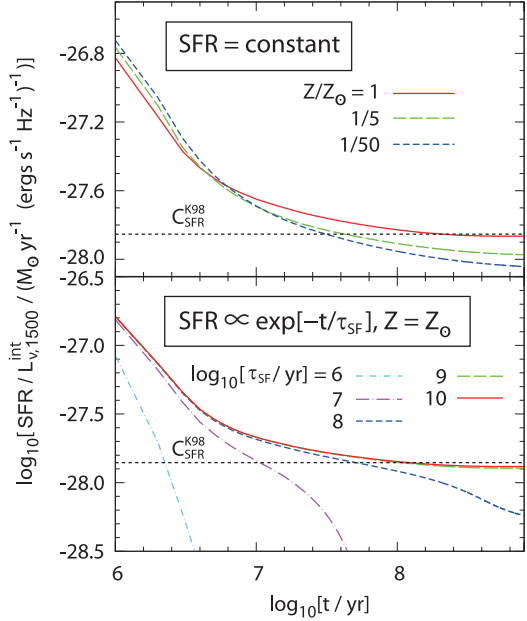}
   \caption{
   Time evolution of $C_\mathrm{SFR}\equiv \mathrm{SFR} / L_{\nu,
   1500}^\mathrm{int}$ calculated using the population synthesis model
   of Schaerer (2003) with a Salpeter IMF in the mass range
   $0.1$--$60~M_\odot$.  Top: $C_\mathrm{SFR}$ for constant star
   formation.  The solid, long-dashed, and short-dotted curves are for
   the stellar metallicities $Z/Z_\odot = 1,\ 1/5,$ and $1/50$,
   respectively.  Bottom: $C_\mathrm{SFR}$ for exponentially decaying
   star formation with an $e$-folding time of $\tau_\mathrm{SF}$ in
   the case of $Z=Z_\odot$.  The solid, long-dashed, short-dashed,
   long-dash-dotted, and short-dash-dotted curves are $C_\mathrm{SFR}$
   at $\log_{10}{\left(\tau_\mathrm{SF} / \mathrm{yr}\right)} = 10$,
   9, 8, 7, and 6, respectively.  The horizontal dotted line in all
   panels represents the most popular conversion factor utilized in
   observational studies given by Kennicutt (1998),
   $C_\mathrm{SFR}^\mathrm{K98} = 1.4 \times 10^{-28}\ M_\odot\
   \mathrm{yr^{-1}\ (erg\ s^{-1}\ Hz^{-1})^{-1}}$.
   }
   \label{fig-TevLlam}
  \end{figure}
  Figure~\ref{fig-TevLlam} shows the time evolution of
  $C_\mathrm{SFR}$ at $\lambda = 1500$~{\AA} for constant and
  exponentially decaying star formation histories (SFHs).  It is
  calculated using the Schaerer (2003) population synthesis model,
  which is used in our model to calculate the UV continuum luminosity
  of galaxies.  As the true SFR of a galaxy having $L_{\nu,
  \mathrm{UV}}^\mathrm{int}$ is represented by $C_\mathrm{SFR} \times
  L_{\nu, \mathrm{UV}}^\mathrm{int}$, the SFR estimated by
  $C_\mathrm{SFR}^\mathrm{K98} \times L_{\nu,
  \mathrm{UV}}^\mathrm{int}$ wrongly represents the true SFR in the
  case of $C_\mathrm{SFR} \neq C_\mathrm{SFR}^\mathrm{K98}$.  When
  $C_\mathrm{SFR}^\mathrm{K98}$ is larger than $C_\mathrm{SFR}$, the
  SFR will be overestimated by a factor of
  $C_\mathrm{SFR}^\mathrm{K98} / C_\mathrm{SFR}$, and vice versa.

  In the early phase of constant star formation, $C_\mathrm{SFR}$ is
  significantly larger than $C_\mathrm{SFR}^\mathrm{K98}$, up to
  $\gtrsim 1$~dex at age $\sim 1$~Myr even in the case of solar
  metallicity $Z_\odot$ as shown in the top panel of
  Figure~\ref{fig-TevLlam}.  This is simply because $L_{\nu,
  \mathrm{UV}}^\mathrm{int}$ grows continuously as the number of
  massive stars with a lifetime of $\sim 100$~Myr increases before it
  reaches equilibrium, which corresponds to
  $C_\mathrm{SFR}^\mathrm{K98}$ for $Z = Z_\odot$.  Therefore,
  adopting $C_\mathrm{SFR}^\mathrm{K98}$ for such young galaxies
  results in an underestimation of SFR for a given $L_{\nu,
  \mathrm{UV}}^\mathrm{int}$.  The extent of the underestimation
  depends on age, metallicity, and wavelength.  Conversely, we
  overestimate the SFR of a galaxy which is old enough to be in
  equilibrium and which has sub-solar metallicity by up to $\sim
  0.2$~dex (a factor of $\sim 1.6$).  This can be understood by the
  fact that lower metallicity stars are bluer; a smaller SFR is enough
  to produce a certain $L_{\nu, \mathrm{UV}}^\mathrm{int}$ compared to
  stars with larger metallicity.  Adopting
  $C_\mathrm{SFR}^\mathrm{K98}$ for such evolved galaxies with
  sub-solar metallicity results in an overestimation of the SFR, the
  extent of the overestimation depending on metallicity and
  wavelength.

  In the case of an exponentially decaying SFH, $C_\mathrm{SFR}$ is
  not constant even at later stages of star formation if the
  $e$-folding time $\tau_\mathrm{SF}$ is short (i.e.,
  $\tau_\mathrm{SF} \lesssim 100$~Myr).  Rather, it decreases
  progressively with time as shown in the bottom panels of
  Figure~\ref{fig-TevLlam}.  This is because $L_{\nu,
  \mathrm{UV}}^\mathrm{int}$ does not decrease as fast with time as
  the SFR.  Massive stars with a lifetime of $\sim 100$~Myr can
  contribute to $L_{\nu, \mathrm{UV}}^\mathrm{int}$ even after the
  star formation activity is quenched in $\tau_\mathrm{SF}$.  The K98
  SFR conversion method for galaxies where SFR decays quickly, like
  starbursts, results in either an underestimation or an
  overestimation of SFR.

  $C_\mathrm{SFR}$ should be smaller than
  $C_\mathrm{SFR}^\mathrm{K98}$ for higher-$z$ galaxies if their UV
  luminosities reach equilibrium because such galaxies typically have
  sub-solar metallicity.  However, if UV LF is dominated by young
  galaxies whose $L_{\nu, \mathrm{UV}}^\mathrm{int}$ is increasing,
  $C_\mathrm{SFR}$ should be larger than
  $C_\mathrm{SFR}^\mathrm{K98}$.  Therefore, it is not straightforward
  to treat $C_\mathrm{SFR}$ for a high-$z$ universe.  In this paper we
  investigate the redshift dependence of $C_\mathrm{SFR}$ in detail
  using the Mitaka model.

  \subsection{Correction of Interstellar Dust Attenuation}
  \label{subsec:CorrDustAtten}

  In order to utilize the UV luminosity of a galaxy as an SFR
  indicator, we need to correct for its interstellar dust attenuation.
  This leads to the most influential uncertainty in the estimation of
  the CSFRD.  This is because the amount of interstellar dust and its
  attenuation are not easily measurable from UV data alone (e.g.,
  Burgarella et al. 2005).  Moreover, the commonly used dust
  attenuation curve rises toward shorter wavelengths.  The required
  luminosity correction can reach $\gtrsim 0.5$~dex at UV wavelengths.

  H04 adopted two independent dust obscuration correction methods for
  the various observed LFs: \textit{common} and \textit{SFR-dependent}
  obscuration corrections.  In both methods, the obscuration
  correction treats the observed LFs as statistical quantities, that
  is, H04 implicitly assumed that the observationally bright galaxies
  are also the intrinsically brightest.

  In the common obscuration correction, H04 adopted $A_V = 0.52$~mag
  for stars; this is a typical obscuration for UV-selected local
  galaxies (see Section 2.2 of H04).  For the dust attenuation curve,
  they adopted the starburst obscuration curve given by Calzetti et
  al. (2000) for all galaxies regardless of their luminosities and
  redshift.  Under these assumptions, the attenuation in magnitude for
  the stellar continuum at a wavelength of 1500~{\AA} is evaluated as
  $A_{1500} = 1.33$~mag.  However, the dust attenuation of high-$z$
  galaxies is not necessarily the same as that of the local
  UV-selected galaxies.

\begin{figure}
   \epsscale{1.15}
   \plotone{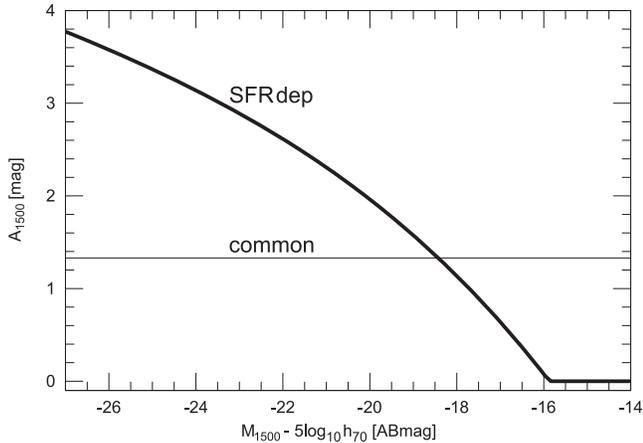}
   \caption{
   Interstellar dust attenuation at $\lambda = 1500$~{\AA},
   $A_{1500}$, as a function of observable (i.e., with dust
   attenuation) absolute magnitude $M_{1500}$ adopted in H04.  The
   thick curve is $A_{1500}$ for the case of the SFR-dependent
   obscuration correction, obtained by solving the transcendental
   Equation~(\ref{eq-Rel-SFRi+SFRo}) numerically.  The thin horizontal
   line represents $A_{1500}$ for the case of the common obscuration
   correction.  $A_{1500}$ at $M_{1500} \gtrsim -16$~mag for the
   SFR-dependent correction is 0 by definition, as in H04.
   }
   \label{fig-MvsAlam}
  \end{figure}

  In the SFR-dependent correction, the obscuration for the UV
  continuum at $\lambda_\mathrm{UV}$, $A_\mathrm{UV}$, of a galaxy
  with observable UV luminosity $L_{\nu, \mathrm{UV}}$ is obtained by
  solving the following transcendental equation numerically:
  \begin{equation}
   A_\mathrm{UV}
    = X(\lambda_\mathrm{UV})
    \log_{10}{\left(\frac{0.797\log_{10}{(L_{\nu,\mathrm{UV}})
	  + 0.318A_\mathrm{UV} + 0.573}}{2.88}\right)};
   \label{eq-Rel-SFRi+SFRo}
  \end{equation}
  where $X(1500~\mbox{\AA}) = 8.935$\footnote{The definition of
  $X(\lambda_\mathrm{UV})$ here differs from that of Hopkins et
  al. (2001) by a factor of 2.5: $X(\lambda_\mathrm{UV}) =
  2.5X^\mathrm{Hop01} (\lambda_\mathrm{UV})$.} in the case of the
  Calzetti attenuation curve.  Figure~\ref{fig-MvsAlam} shows
  $A_\mathrm{UV}$ for both the SFR-dependent correction and the common
  correction as a function of observable absolute magnitude at
  $\lambda_\mathrm{UV} = 1500$~{\AA}.
  As shown in Figure~\ref{fig-MvsAlam}, H04 adopted $A_{1500} = 0$~mag
  at $M_{1500} \gtrsim -16$~mag because the numerical solution of
  Equation~(\ref{eq-Rel-SFRi+SFRo}) for $A_\mathrm{UV}$ becomes
  negative and this is physically meaningless.

  Equation~(\ref{eq-Rel-SFRi+SFRo}) was originally derived from an
  empirical relation between $E(B-V)_\mathrm{gas}$ and the far-IR
  (FIR) luminosity $L_\mathrm{FIR}$ for normal galaxies, blue compact
  galaxies, and starbursts at the local universe, in which brighter
  galaxies have larger obscuration (see references given in Section
  2.2 of H04).  However, the distribution of these galaxies in the
  $E(B-V)_\mathrm{gas}$--$L_\mathrm{FIR}$ plane has significant
  scatter around the empirical relation (see Figure 1 of Hopkins et
  al. 2001).  Moreover, the numerical constant in the numerator of the
  right-hand side in Equation~(\ref{eq-Rel-SFRi+SFRo}) contains the
  uncertainty in $\mathrm{SFR}/L_\mathrm{FIR}$ and $\mathrm{SFR} /
  L_{\nu, \mathrm{UV}}^\mathrm{int}$.  Even if this obscuration
  correction method had been derived empirically in the local
  universe, we should examine its validity in the high-$z$ universe.
  Thus, we investigate here the redshift dependence of the mean of
  $A_\mathrm{UV}$ in detail.

  HB06 adopted a different approach from to of H04 at $z < 3$; they
  added the FIR measurements of the CSFRD to the
  obscuration-uncorrected UV data as an effective dust obscuration
  correction in order to avoid assumptions about the extent or form of
  the obscuration and variations due to possible luminosity bias in
  the UV-selected sample (see Section 2.2 of HB06).  At $z>3$, where
  there is no reliable FIR measurement of the CSFRD, HB06 adopted the
  common obscuration correction as in H04.  As we do not calculate IR
  emission in this paper, we cannot examine the validity of the HB06
  method.  However, we give a brief comment on the consistency with
  the HB06 CSFRD in Section \ref{subsec:Comments2HB06}.
 \section{Model Description}\label{sec:Model}

 In order to examine the cause of the disagreement between the CSFRDs
 obtained in theoretical and observational studies, we utilize a mock
 catalog of galaxies generated by a semi-analytic model for galaxy
 formation called the Mitaka model (Kobayashi et al. 2007, 2010;
 updated version of Nagashima \& Yoshii 2004).  The Mitaka model
 follows the framework of the $\Lambda$CDM model of structure
 formation and calculates the redshift evolution of the physical
 quantities of each galaxy at any redshift via semi-analytic
 computation of the merger histories of dark matter halos and the
 evolution of baryon components within halos.  The time evolution of
 baryon components within halos is followed using physically motivated
 phenomenological models for radiative cooling, star formation,
 supernova feedback, chemical enrichment of gas and stars, and galaxy
 mergers.  As a result, various physical and observational properties
 such as the SFR and intrinsic and observable stellar continuum
 luminosities of the galaxies at any given redshift can be obtained.

 A detailed description of our model is given in Nagashima \& Yoshii
 (2004) and Kobayashi et al. (2007, 2010).  Therefore, here we briefly
 describe some key prescriptions which are closely related to this
 paper.  Then, we explain how to examine the origin of the difference
 in CSFRD between observation and theory by using the mock catalog of
 galaxies generated by the Mitaka model.

  \subsection{Merger Tree of Dark Matter Halos}
  \label{subsec-mergertree}

  The merger histories of dark matter halos are realized using a Monte
  Carlo method based on the extended Press-Schechter formalism (Bond
  et al. 1991; Bower 1991; Lacey \& Cole 1993).  For the halo mass
  function to provide the weight for summing merger trees, the Mitaka
  model adopts the analytic functional form given by Yahagi et
  al. (2004); this is a fitting function to their high-resolution
  $N$-body simulation.

  \begin{figure}
   \epsscale{1.17}
   \plotone{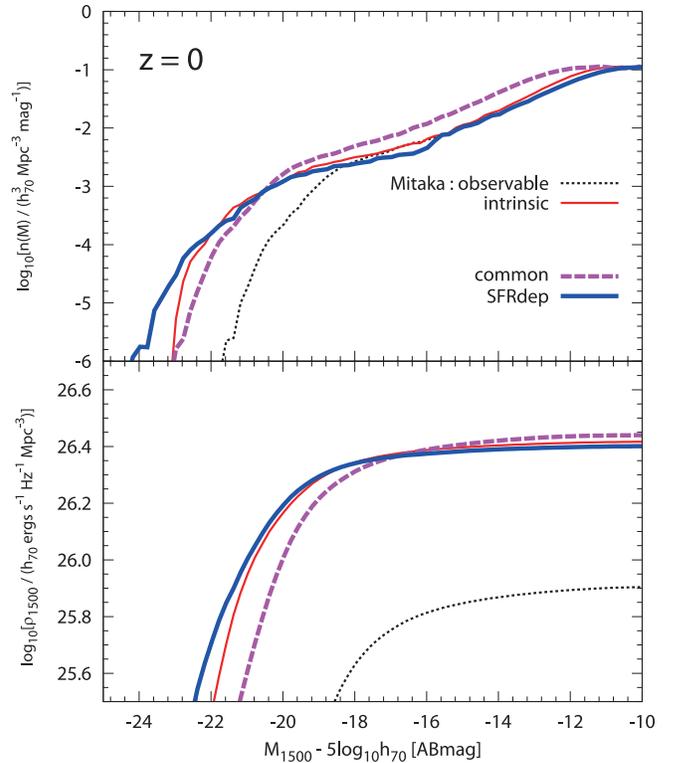}
   \caption{
   1500~{\AA} LFs (top) and LDs (bottom) as a function of absolute
   magnitude and limiting absolute magnitude for integration,
   respectively, at $z=0$.  The thin solid and dotted curves are the
   intrinsic and observable quantities of the Mitaka model,
   respectively.  The thick solid and dashed curves are the common and
   SFRdep LFs and LDs calculated by applying the common and
   SFR-dependent obscuration corrections of H04 to the observable LF
   of the Mitaka model.
   }
   \label{fig-UVLF+UVLD-z0}
  \end{figure}
  Dark matter halos with circular velocity $V_\mathrm{c} \ge
  V_\mathrm{low} = 30\ \mathrm{km\ s^{-1}}$ are regarded as isolated
  halos, corresponding to the lower limits of halo mass
  $M_\mathrm{halo} \gtrsim 10^9\ M_\odot$ and $10^{10}\ M_\odot$ at
  $z=5$ and 0, respectively.  Because of the existence of
  $V_\mathrm{low}$, fainter galaxies (i.e., $M_{1500} \gtrsim
  -10$~mag) are not well resolved in the Mitaka model.  However, this
  limited resolution does not affect the resulting quantities at
  1500~{\AA}, $\rho_\mathrm{1500}$ and $\dot{\rho}_\star$, because
  $M_{1500} \sim -10$~mag is much fainter than $M_\ast$ and
  $\rho_\mathrm{1500}$ converges well at magnitudes much brighter
  those shown in the bottom panel of Figure~\ref{fig-UVLF+UVLD-z0}.

  \newpage
  \subsection{Stellar Continuum Luminosity}

  The intrinsic luminosities and colors of galaxies in our model are
  calculated according to their SFHs and chemical enrichment histories
  using a stellar population synthesis model.  We emphasize that we do
  not assume a simple linear relation between the intrinsic luminosity
  and SFR even at UV wavelengths.  While the original Mitaka model
  uses the population synthesis model of Kodama \& Arimoto (1997), our
  model uses that of Schaerer (2003) to calculate the UV stellar
  continuum luminosity as in Kobayashi et al. (2007, 2010).  Compared
  to the Kodama \& Arimoto (1997) model, the Schaerer (2003) model is
  a more recent one and covers a wider range of metallicity (including
  zero metallicity).

  It has already been shown (Kobayashi et al. 2007, 2010) that the
  revised version of the Mitaka model reproduces all well the observed
  statistical quantities of high-$z$ Ly$\alpha$ emitters and
  Lyman-break galaxies.

  \subsection{Prescription for Dust Attenuation}
  \label{subsec-MitakaDust}

  In order to calculate observational (i.e., with interstellar dust
  attenuation) luminosities and colors of model galaxies, we model the
  optical depth of their internal dust as follows.  We make the usual
  assumption that dust abundance of a galaxy is proportional to its
  gas metallicity and that the dust optical depth $\tau_\mathrm{d}$ is
  proportional to the column density of metals:
  \begin{equation}
   \tau_\mathrm{d} \propto
    \frac{M_\mathrm{c} Z_\mathrm{c}}{r_\mathrm{e}^2},
    \label{eq-taud-mitaka}
  \end{equation}
  where $M_\mathrm{c}$ and $Z_\mathrm{c}$ are the mass and metallicity
  of cold gas, respectively, and $r_\mathrm{e}$ is the effective
  radius, all of which are obtained in our model.  The normalization
  of Equation~(\ref{eq-taud-mitaka}) is assumed to be a constant and
  universal regardless of galaxy properties and redshift, and has been
  determined to fit the observed data of local galaxies.

  The wavelength dependence of the dust optical depth
  $\tau_\mathrm{d}(\lambda)$ is adopted to be the same as the Galactic
  extinction curve given by Pei (1992).  In terms of the dust
  distribution, our Mitaka model assumes slab geometry (i.e., a
  uniform distribution of sources and absorbers; see Equation~(19) of
  Calzetti et al. 1994 or Section 3.2 of Clemens \& Alexander 2004).
  Therefore, the amount of attenuation magnitude $A_\lambda$ for
  the stellar continuum is given by
  \begin{equation}
   10^{-0.4A_\lambda} =
    \frac{1-\exp{(-\tau_\mathrm{d}(\lambda))}}{\tau_\mathrm{d}(\lambda)}.
    \label{eq-Alam}
  \end{equation}
  Note that, while the wavelength dependence of $A_\lambda$ calculated
  via Equation~(\ref{eq-Alam}) is similar to that of $\tau_\mathrm{d}$
  (i.e., the Galactic extinction curve) at $\tau_\mathrm{d} \ll 1$,
  they are significantly different at $\tau_\mathrm{d} \gg 1$:
  $A_\lambda$ becomes flatter than $\tau_\mathrm{d}$ at
  $\tau_\mathrm{d} \gg 1$ as $A_\lambda \propto \log_{10}
  {\tau_\lambda}$.  Therefore, the wavelength dependence of
  $A_\lambda$ in our model at $\tau_\mathrm{d} \gg 1$ is significantly
  different from that of the Calzetti law.

  We emphasize that our model calculates star formation and chemical
  enrichment consistently in the framework of hierarchical structure
  formation.  Therefore, both the time evolution of $\tau_\mathrm{d}$
  for each model galaxy and the redshift evolution of its mean at a
  certain redshift can be naturally incorporated into our model
  together with the evolution of $M_\mathrm{c}$, $Z_\mathrm{c}$ and
  $r_\mathrm{e}$.

  \subsection{Application to the CSFRD Study}

  In this paper, we calculate the CSFRD in the redshift range
  $z=0$--10 using the mock catalog of galaxies generated by our model
  following the prescriptions in H04.  We treat the mock catalog like
  observational data compiled from the literature and utilized to
  obtain the CSFRD in H04.  We emphasize the most striking difference
  between our catalog and the data compiled in H04 is that each model
  galaxy in our catalog has information on its intrinsic UV luminosity
  and SFR as well as the observable luminosity.

  We apply the same obscuration corrections as H04 (i.e., common and
  SFR-dependent corrections) to the observable UV LFs at $z=0$--10
  given by our model in order to obtain the intrinsic LFs, which are
  represented as \textit{common} and \textit{SFRdep LFs},
  respectively.  Their LDs are represented as
  $\rho_\mathrm{UV}^\mathrm{com}$ and
  $\rho_\mathrm{UV}^\mathrm{SFRdep}$, respectively.  We note here that
  the observable UV LF given by our model can be separated into
  contributions from each model galaxy.  Therefore, the intrinsic UV
  LF can be obtained directly by counting the contributions of model
  galaxies in each magnitude bin.  However, we treat the observable UV
  LF as a statistical quantity to examine whether the H04 obscuration
  corrections reproduce the ``true'' intrinsic UV LF and
  $\rho_\mathrm{UV}^\mathrm{int}$.

  The conversion factor $C_\mathrm{SFR}^\mathrm{K98}$ is also compared
  with that of our model both in each galaxy, $\mathrm{SFR} / L_{\nu,
  \mathrm{UV}}^\mathrm{int}$, and in all galaxies, $\dot{\rho}_\star /
  \rho_\mathrm{UV}^\mathrm{int}$.  We note again that the intrinsic UV
  luminosity of a galaxy is calculated according to its detailed SFH
  and chemical enrichment history; therefore $C_\mathrm{SFR}$ can vary
  in every galaxy.

  We note here that, while there are several free parameters in
  phenomenological models for baryon evolution, in our model, they
  have already been determined to fit the various observations of the
  local galaxies (i.e., \textit{B}- and \textit{K}-band LFs, neutral
  gas mass fraction, and the gas mass-to-luminosity ratio as a
  function of \textit{B}-band luminosity) in Nagashima \& Yoshii
  (2004).  As we utilize the values from this work unchanged in our
  study, there are no free parameters that can be adjusted.

 \section{Comparison with H04 Prescriptions}
 \label{sec:Results}

  \subsection{Dust Obscuration Correction}
  \label{subsec:result-dust}

  We first show the results from the H04 approaches to correcting for
  interstellar dust obscuration.  The intrinsic and observable UV LFs
  at $1500$~{\AA} at $z=0$ are shown in the top panel of
  Figure~\ref{fig-UVLF+UVLD-z0}.

  The common and SFRdep LFs (top panel) and their LDs (bottom panel)
  are also shown.  Both LDs agree with the intrinsic one if we
  integrate the LFs up to $\gtrsim -16$~mag, even though both LFs show
  a significant deviation from the intrinsic one.  This might be
  trivial because both of the H04 corrections are calibrated by
  galaxies in the local universe.

  \begin{figure}
   \epsscale{1.2}
   \plotone{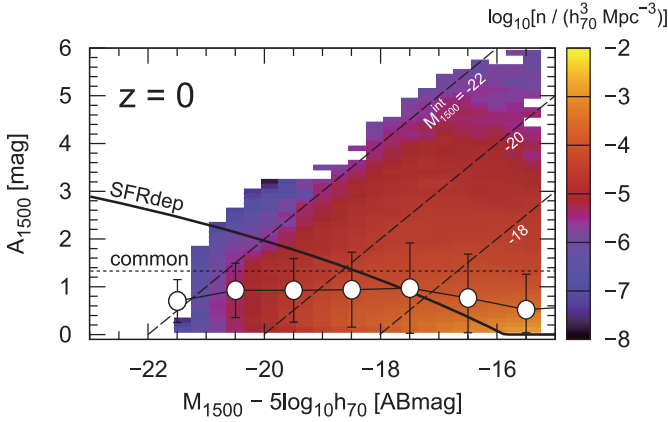}
   \caption{
   Distribution of model galaxies with $M_{1500}^\mathrm{int} \leq
   -15$~mag at $z=0$ in the $A_{1500}$--$M_{1500}$ plane.  The color
   scale shows the color-coded values of the number density of model
   galaxies per unit area.  The open circles with error bars connected
   by lines are the mean and $1\sigma$ for the model galaxies in each
   magnitude bin.  The obscuration corrections of H04 are also shown
   as the solid curve and dotted line for the SFR dependent and common
   corrections, respectively.  The dashed lines indicate the contours
   of $M_{1500}^\mathrm{int}$ corresponding to $-22$, $-20$, and
   $-18$~AB mag from left to right.
   }
   \label{fig-Mag-Alam-z0}
  \end{figure}
  In order to see the origin of the similarities and discrepancies
  between the UV LF and LD given by our model and those using the H04
  methods, we compare the dust attenuation of model galaxies as a
  function of absolute magnitude $M_{1500}$ with
  $A_{1500}^\mathrm{com}$ and $A_{1500}^\mathrm{SFRdep}$.  As shown in
  Figure~\ref{fig-Mag-Alam-z0}, the means of $A_{1500}$, $\langle
  A_{1500} \rangle$, for our model galaxies at each magnitude are
  found to be almost constant, $\langle A_{1500} \rangle \approx
  0.6$--1.0~mag, for $-22~\mathrm{mag} \lesssim M_{1500} \lesssim
  -15$~mag.  However, model galaxies with a certain magnitude have
  significant dispersion around $\langle A_{1500} \rangle$ up to $\sim
  1.0$~mag.  This means that a non-negligible number of intrinsically
  bright galaxies lies at observationally faint magnitudes.  The
  contribution from such observationally faint and intrinsically
  bright galaxies to the intrinsic LF and LD is completely neglected
  if a single value of $\langle A_{1500} \rangle$ at a certain
  magnitude is adopted as the representative attenuation for all
  galaxies with this magnitude.  Therefore, correcting for dust
  attenuation by using $\langle A_{1500} \rangle$ results in
  underestimation of the bright ends of the intrinsic LF and intrinsic
  LD.

  Taking the consideration above into account, we compare the H04
  obscuration corrections with the median quantity $\langle A_{1500}
  \rangle$ for the model galaxies.  The attenuation in the common
  correction of $A_{1500}^\mathrm{com} = 1.33$~mag is found to be
  larger than $\langle A_{1500} \rangle$ by $\approx 0.3$--0.7~mag,
  that is, within the 1$\sigma$ error except for the brightest
  magnitudes.  As shown in the top panel of
  Figure~\ref{fig-UVLF+UVLD-z0}, this overcorrection for dust
  attenuation compared to the correction with $\langle A_{1500}
  \rangle$ is not sufficient to compensate for the contribution of the
  observationally faint and intrinsically bright galaxies at the
  bright end of the intrinsic LF.  However, the overcorrection at
  faint magnitudes is enough to cover the underestimation of the LD at
  the bright end.  Hence, the common correction results in a similar
  intrinsic UV LD to that of the Mitaka model.  On the other hand,
  attenuation in the SFR-dependent correction of
  $A_{1500}^\mathrm{SFRdep}$ is significantly different from $\langle
  A_{1500} \rangle$.  We note that it is also different from the
  recent observational result for star-forming galaxies with $z =
  0.95$--2.2, which shows that the mean attenuation at far-UV
  wavelengths decreases toward the bright far-UV continuum luminosity
  (see Figure~7 of Buat et al. 2012).  It is larger and smaller than
  $\langle A_{1500} \rangle$ in the magnitude range $M_{1500} \lesssim
  -18$~mag and $M_{1500} \gtrsim -18$~mag, respectively.  The
  overcorrection at bright magnitudes leads to the overestimation of
  the bright end of the intrinsic LF as shown in
  Figure~\ref{fig-UVLF+UVLD-z0}.  It is compensated for by the
  underestimation of the number density of faint galaxies.  As a
  result, the SFR-dependent correction also results in a similar
  intrinsic UV LD to that of the Mitaka model.

  \begin{figure*}
   \epsscale{1.}
   \plotone{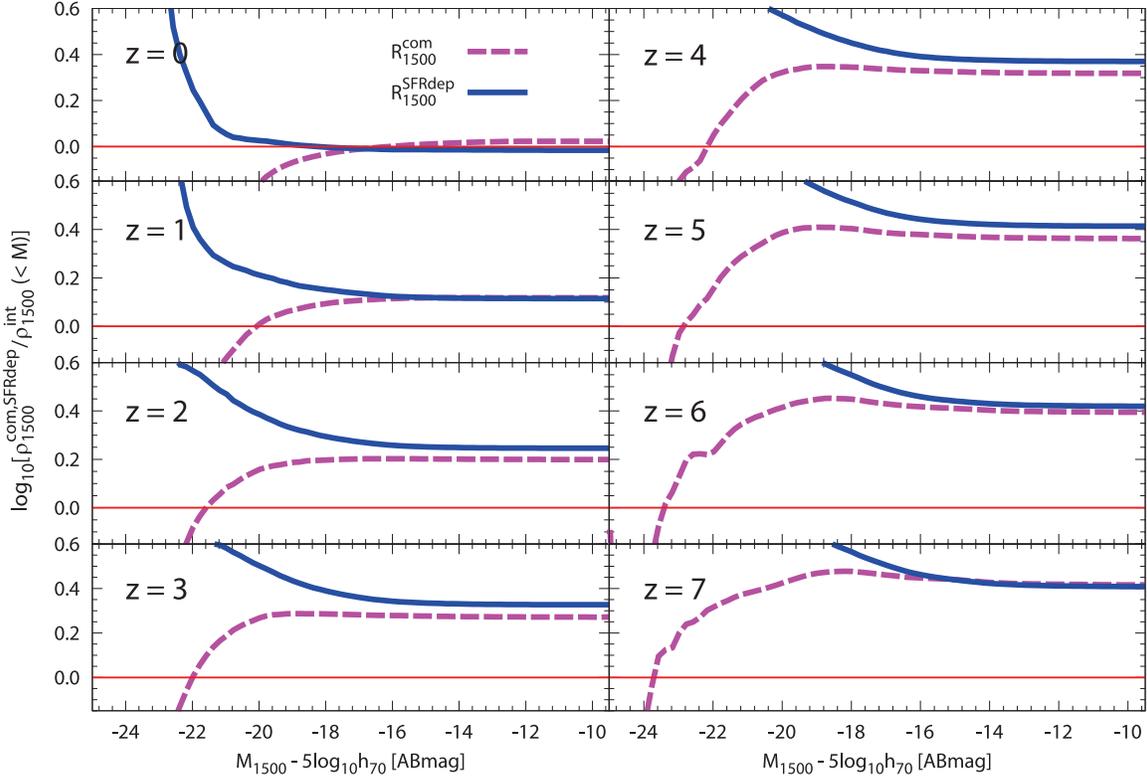}
   \caption{
   LD ratios $R_{1500}^\mathrm{com, SFRdep} \equiv
   \rho_{1500}^\mathrm{com, SFRdep} / \rho_{1500}^\mathrm{int}$ as a
   function of absolute magnitude for the wavelength $\lambda =
   1500$~{\AA} in the redshift range $z=0$--7 from top left to bottom
   right.  The line styles are the same as in
   Figure~\ref{fig-UVLF+UVLD-z0}.
   }
   \label{fig-Ratio-z0-3}
  \end{figure*}
  We show the redshift evolution of the intrinsic 1500~{\AA} LD ratios
  $R_{1500}^i \equiv \rho_{1500}^i / \rho_{1500}^\mathrm{int}$, where
  $i = \mathrm{com}$ or SFRdep, as a function of magnitude in
  Figure~\ref{fig-Ratio-z0-3}.  As both correction methods ignore the
  contribution of observationally faint and intrinsically bright
  galaxies, the values of $R_{1500}^i$ depart significantly from unity
  at bright magnitudes.  However, at faint magnitudes the $R_{1500}^i$
  rapidly converge to the same value; the asymptotic values of
  $R_{1500}^i$ at $z=0$ are found to be unity, as seen in
  Figure~\ref{fig-UVLF+UVLD-z0}.  However, it is easily seen that the
  asymptotic values increase with $z$ and reach $\sim 0.3$~dex and
  $\sim 0.4$~dex at $z=3$ and $z\ge 4$, respectively.  In summary, the
  H04 obscuration corrections alone result in the overestimation of
  CSFRD by $\sim 0.1$--0.4~dex in the redshift range $z=0$--10
  compared to that of our model galaxies.

  \subsection{Ratio of UV Luminosity to SFR}
  \label{subsec-ResCSFR}

\begin{figure}
   \epsscale{1.2}
   \plotone{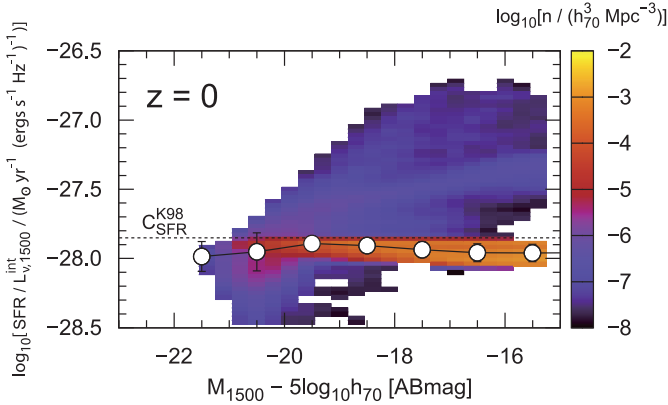}
   \caption{
   Same as Figure~\ref{fig-Mag-Alam-z0}, but for the distribution in
   the $C_\mathrm{SFR}$--$M_{1500}$ plane.
   $C_\mathrm{SFR}^\mathrm{K98} = 1.4\times 10^{-28}\ M_\odot\
   \mathrm{yr^{-1}\ (erg\ s^{-1}\ Hz^{-1})^{-1}}$ is also shown as a
   dotted horizontal.
   }
   \label{fig-Mag-CSFR-z0}
  \end{figure}

  In Figure~\ref{fig-Mag-CSFR-z0}, we show the distribution of model
  galaxies with $M_{1500}^\mathrm{int} \leq -15$~mag at $z=0$ in the
  $C_\mathrm{SFR}$--$M_{1500}$ plane.  It is found there are two
  sequences in the plane: a constant sequence at $\sim
  C_\mathrm{SFR}^\mathrm{K98}$ and a widely spread sequence around the
  constant sequence with relatively high and low number densities
  ($\gtrsim 10^{-4}~h_{70}^3~\mathrm{Mpc^{-3}}$ and $\lesssim
  10^{-6}~h_{70}^3~\mathrm{Mpc^{-3}}$), respectively.  These sequences
  correspond to the two distinctive galaxy populations in our model,
  that is, quiescently star-forming and starburst galaxies,
  respectively.

  The quiescent galaxies have long star formation timescales ($\gtrsim
  1$~Gyr) and form disk stars almost constantly in the timestep of the
  Mitaka model.  Therefore, the quantities contributing to
  $C_\mathrm{SFR}$ already reach equilibrium, with values depending on
  the stellar metallicity.  As there is a well-known
  magnitude--metallicity relation (i.e., fainter galaxies have smaller
  metallicity; e.g., Garnett 2002), $C_\mathrm{SFR}$ decreases toward
  fainter magnitudes.  In contrast, the starburst galaxies, whose
  starburst activity is triggered by a major merger of galaxies, have
  short star formation timescales ($\tau_\mathrm{SF} \lesssim 1$~Gyr)
  determined via the dynamical timescale of a newly formed spheroid.
  Because the starburst onset times are chosen randomly in the
  timestep of the Mitaka model, they have a variety of
  $C_\mathrm{SFR}$s according to the onset time and $\tau_\mathrm{SF}$
  as shown in the bottom panel of Figure~\ref{fig-TevLlam}.  The
  starburst galaxies whose $C_\mathrm{SFR}$ are larger than
  $C_\mathrm{SFR}^\mathrm{K98}$ are relatively young and those with
  $C_\mathrm{SFR} < C_\mathrm{SFR}^\mathrm{K98}$ are old enough to
  have such small $C_\mathrm{SFR}$.

  We also evaluate the mean $C_\mathrm{SFR}$, $\langle C_\mathrm{SFR}
  \rangle$, for all our model galaxies at each magnitude.  Starburst
  galaxies dominate the bright end in $M_{1500}$ at $z=0$ ($\lesssim
  -20$~mag) while quiescent galaxies determine $\langle C_\mathrm{SFR}
  \rangle$ except for the bright end. As a combination of the
  contributions from these two distinctive populations, $\langle
  C_\mathrm{SFR} \rangle$ is found to be almost constant in all of the
  magnitude range shown in Figure~\ref{fig-Mag-CSFR-z0} and is always
  smaller than $C_\mathrm{SFR}^\mathrm{K98}$.  Therefore, converting
  $L_{\nu,\mathrm{UV}}^\mathrm{int}$ into SFR via
  $C_\mathrm{SFR}^\mathrm{K98}$ results in an overestimation of the
  SFR for most galaxies at $z=0$, although the overestimation is not
  significant ($\lesssim 0.1$~dex).

\begin{figure*}
   \epsscale{0.75}
   \plotone{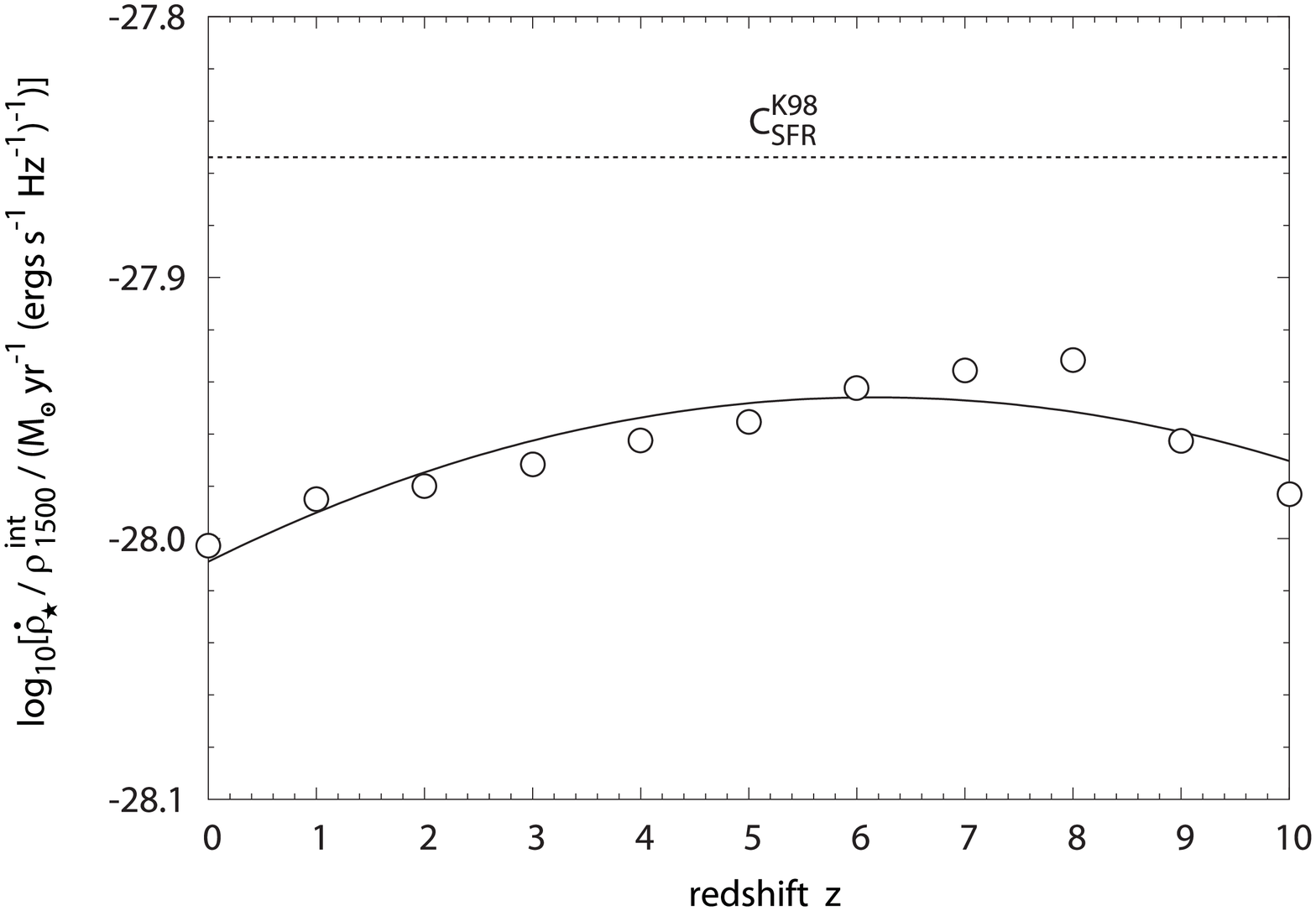}
   \caption{
   Ratio of the CSFRD and intrinsic 1500~{\AA} LD (i.e.,
   $C_\mathrm{SFR}^\mathrm{eff}$) of the Mitaka model as a function of
   redshift.  The open circles are the model results and the solid
   curve shows the best-fit quadratic function defined by
   Equation~(\ref{eq-Lnu2SFR}); the best-fit parameters are compiled
   in Table~\ref{tab-FitParameters}.  Note that the dynamic range of
   the vertical axis (0.3~dex) is much smaller than that of
   Figure~\ref{fig-Mag-CSFR-z0} (2.0~dex).
   }
   \label{fig-CSFReff-z}
  \end{figure*}

  Let us define the \textit{effective} $L_{\nu,
  \mathrm{UV}}^\mathrm{int}$-to-SFR conversion factor,
  $C_\mathrm{SFR}^\mathrm{eff}(z)$, via:
  \begin{equation}
   C_\mathrm{SFR}^\mathrm{eff}(z) \equiv \dot{\rho}_\star(z) /
    \rho_\mathrm{UV}^\mathrm{int}(z).
  \end{equation}
  This is \textit{not} a conversion factor for each galaxy
  \textit{but} by definition is the direct conversion from the
  statistical quantity of the intrinsic UV LD,
  $\rho_\mathrm{UV}^\mathrm{int}$, into the CSFRD, $\dot{\rho_\star}$.
  $\dot{\rho}_\star$ and $\rho_\mathrm{UV}^\mathrm{int}$ can be
  rewritten as summations of the contribution from each galaxy via
  $\dot{\rho}_\star = \sum \mathrm{SFR}^i = \sum ( C_\mathrm{SFR}^i
  \times L_{\nu,1500}^{\mathrm{int},i} )$ and
  $\rho_\mathrm{UV}^\mathrm{int} = \sum
  L_{\nu,1500}^{\mathrm{int},i}$, where the suffix $i$ indicates a
  galaxy.  Hence, $C_\mathrm{SFR}^\mathrm{eff} = \sum (
  C_\mathrm{SFR}^i \times L_{\nu, 1500}^{\mathrm{int},i} ) / \sum
  L_{\nu, 1500}^{\mathrm{int},i}$ also represents a mean conversion
  factor of $C_\mathrm{SFR}$ weighted by the intrinsic UV continuum
  luminosity $L_{\nu,1500}^\mathrm{int}$.  Figure~\ref{fig-CSFReff-z}
  shows the redshift evolution of $C_\mathrm{SFR}^\mathrm{eff}(z)$ at
  $\lambda = 1500$~{\AA}.
  As shown in Figure~\ref{fig-CSFReff-z},
  $C_\mathrm{SFR}^\mathrm{eff}$ at $z=0$ is smaller than
  $C_\mathrm{SFR}^\mathrm{K98}$ by $\sim 0.15$~dex.  This result is
  natural because $\langle C_\mathrm{SFR} \rangle$ is smaller than
  $C_\mathrm{SFR}^\mathrm{K98}$ over the magnitude range $-22\
  \mathrm{mag} \lesssim M_{1500} \lesssim -15$~mag as shown in
  Figure~\ref{fig-Mag-CSFR-z0}.  This implies that converting
  $\rho_\mathrm{UV}^\mathrm{int}$ for the model galaxies at $z=0$ into
  $\dot{\rho}_\star$ using $C_\mathrm{SFR}^\mathrm{K98}$ overestimates
  $\dot{\rho}_\star$ compared to its true quantity by $\approx
  0.15$~dex.  The redshift evolution of $C_\mathrm{SFR}^\mathrm{eff}$
  is also shown in Figure~\ref{fig-CSFReff-z}.  It cannot be not
  simply understood because of the following effects.  Toward high
  redshift, the contribution from young starbursts increases because
  of the increasing rate of the major merger.  This effect results in
  an increase in $C_\mathrm{SFR}^\mathrm{eff}$.  At the same time, the
  dynamical timescale of starburst galaxies becomes shorter at higher
  redshift because the galaxies formed tend to be smaller in size and
  hence have shorter $\tau_\mathrm{SF}$.  This effect results in a
  decrease in $C_\mathrm{SFR}^\mathrm{eff}$ because $C_\mathrm{SFR}$
  rapidly decays in galaxies with short $\tau_\mathrm{SF}$.  These two
  competing effects lead to an increase in
  $C_\mathrm{SFR}^\mathrm{eff}$ toward $z\lesssim 8$ and its decrease
  at $z\gtrsim 8$.

  $C_\mathrm{SFR}^\mathrm{eff}$ for $\lambda = 1500$~{\AA} is found to
  be always smaller than $C_\mathrm{SFR}^\mathrm{K98}$ by $\approx
  0.10$--0.15~dex.  The reason why $C_\mathrm{SFR}^\mathrm{eff}$ is
  always smaller than $C_\mathrm{SFR}^\mathrm{K98}$ is as follows.  As
  shown in Figure~\ref{fig-TevLlam}, the time duration during which
  the galaxies have $C_\mathrm{SFR}$ larger than
  $C_\mathrm{SFR}^\mathrm{K98}$ is short ($\lesssim 100$~Myr) for any
  metallicity and for both constant and exponentially declining SFHs.
  This implies that the chance probability to detect a galaxy with
  $C_\mathrm{SFR} > C_\mathrm{SFR}^\mathrm{K98}$ is smaller than the
  probability to observe a galaxy with $C_\mathrm{SFR} <
  C_\mathrm{SFR}^\mathrm{K98}$.  Moreover, high-$z$ galaxies tend to
  have a metallicity smaller than $Z_\odot$ and hence the equilibrium
  value of $C_\mathrm{SFR}$ is smaller than
  $C_\mathrm{SFR}^\mathrm{K98}$ for the constantly star-forming
  galaxies.  As a consequence of these two reasons, the mean of
  $C_\mathrm{SFR}$ becomes smaller than $C_\mathrm{SFR}^\mathrm{K98}$
  at all redshifts.

  In summary, adopting the usual conversion factor of
  $C_\mathrm{SFR}^\mathrm{K98}$ for galaxies at $z=0$--10 results in
  an overestimation of $\dot{\rho}_\star$ by $\approx 0.10$--0.15~dex.
 \section{Summary and Discussion}\label{sec:Summary+Discussion}

 In this paper, we examined the cause of the inconsistency in the
 CSFRD between theoretical and observational studies, focusing on an
 SFR indicator for a high-$z$ universe, that is, the rest-frame
 1500~{\AA} stellar continuum luminosity $L_{1500}$.  By using a
 semi-analytic model for galaxy formation, the so-called Mitaka model,
 we found that the underestimation of the CSFRD seen in theoretical
 model originates from the following two uncertainties in the process
 to evaluate the CSFRD from the observed 1500~{\AA} LD: the dust
 obscuration correction and the conversion from $L_{1500}$ to SFR.
 The uncertainty in the faint-end slope of the LF is not the origin of
 the underestimation of CSFRD but the origin of the dispersion around
 the median CSFRD.  The methods of obscuration correction adopted in
 H04 result in the overestimation of the CSFRD by $\approx
 0.1$--0.4~dex and the SFR conversion used in observational studies
 also leads to the overestimation of the CSFRD by $\approx
 0.1$--0.2~dex.

 Since theoretical models including ours reproduce the observed data
 for the UV LD which is not corrected by dust attenuation and the data
 for the SMD, the inconsistency in the CSFRD does not imply that
 theoretical models miss some key physical processes in galaxy
 formation.  Of course, the theoretical models are not yet perfect
 because observed data such as cosmic downsizing (e.g., Cowie et
 al. 1996) remain to be reproduced.  The revision of theoretical
 models can be achieved through a comparison with direct observed data
 which are \textit{not} affected by a certain model and/or
 assumptions.

 In this section, we provide a brief discussion of the origin of the
 difference in dust attenuation at high $z$ between our model and the
 H04 corrections.  We also present new empirical calibrations for dust
 attenuation and SFR conversion as well as a recipe for utilizing them
 in observational studies.

  \subsection{Origin of the Difference in Dust Attenuation at High
  Redshift}

  As described in Section~\ref{subsec:result-dust}, the H04
  obscuration correction methods reproduce the intrinsic LD from the
  observable UV LF of our model galaxies at $z=0$, although they
  overestimate it at higher redshifts.  Here we discuss the origin of
  the overestimation at high redshift.

\begin{figure}
   \epsscale{1.15}
   \plotone{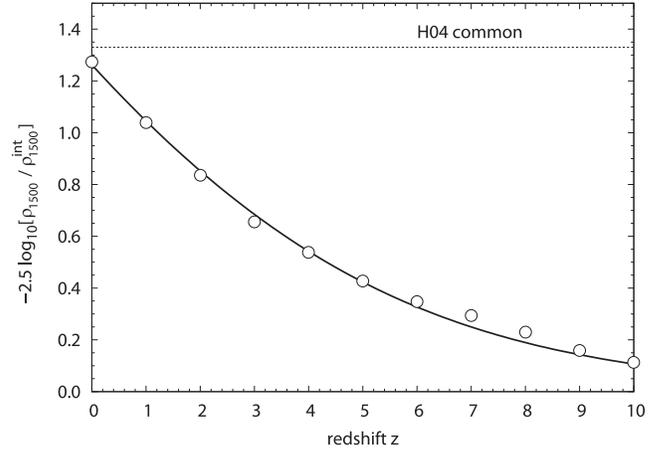}
   \caption{
   Same as Figure~\ref{fig-CSFReff-z}, but for the ratio of the
   observable 1500~{\AA} LD to the intrinsic one (i.e.,
   $A_{1500}^\mathrm{eff}$) of the Mitaka model as a function of
   redshift.  The parameters for the best-fit analytic function
   defined by Equation~(\ref{eq-CdustEff}) and represented as the
   thick solid curve are also given in Table~\ref{tab-FitParameters}.
   }
   \label{fig-Alameff-z}
  \end{figure}

  Our model naturally incorporates the redshift evolution of dust
  abundance because it calculates chemical enrichment of gas in each
  model galaxy consistently according to its SFH.  This can be seen in
  Figure~\ref{fig-Alameff-z}, which shows the redshift evolution of
  \textit{effective} dust attenuation in magnitude,
  $A_{1500}^\mathrm{eff}$, for our model galaxies defined by
  \begin{equation}
   A_{1500}^\mathrm{eff}(z) \equiv -2.5
    \log_{10}{\left(\rho_{1500}(z) /
	  \rho_{1500}^\mathrm{int}(z) \right)}.
    \label{eq-Alameff}
  \end{equation}

  As $A_{1500}^\mathrm{eff}$ can be rewritten using the $A_{1500}$ and
  $L_{\nu,1500}^\mathrm{int}$ for each galaxy as
  $A_{1500}^\mathrm{eff} = -2.5\log_{10} ( \sum ( 10^{-0.4A_{1500}^i}
  \times L_{\nu, 1500}^{\mathrm{int}, i} ) / $ $\sum
  L_{\nu,1500}^{\mathrm{int},i})$, $A_{1500}^\mathrm{eff}$ represents
  the mean dust attenuation weighted by the intrinsic UV continuum
  luminosity $L_{\nu, 1500}^\mathrm{int}$.  This is the reason why
  $A_{1500}^\mathrm{eff}$ at $z=0$ ($= 1.3$~mag) is larger than
  $\langle A_{1500} \rangle$ at $z=0$ ($\lesssim 1$~mag).
  $A_{1500}^\mathrm{eff}$ becomes smaller at higher $z$ because of the
  redshift evolution of metallicity and dust abundance.  However, such
  a redshift evolution of dust abundance is not incorporated into the
  H04 obscuration corrections.  This is the origin of their
  overestimation of the intrinsic LD.

  A quantity similar to $A_{1500}^\mathrm{eff}$ has been evaluated
  from the observed LD ratio between IR and UV, $\rho_\mathrm{IR} /
  \rho_\mathrm{UV}$ ($\approx \rho_{1500}^\mathrm{int} / \rho_{1500} -
  1$), in the redshift range $z=0$--1 (Takeuchi et al. 2005).  They
  found that $A_{1500}^\mathrm{eff}$ increases monotonically toward
  $z=1$, which is the opposite trend to our result.  However, more
  recent observational estimates for $\rho_\mathrm{IR}$ (e.g., Murphy
  et al. 2011; Casey et al. 2012; Cucciati et al. 2012) find that the
  redshift evolution of $\rho_\mathrm{IR}$ in this redshift range is
  milder than that reported by Takeuchi et al. (2005).  Cucciati et
  al. (2012) also find that mean dust attenuation decreases toward
  high $z$ at $z\gtrsim 1$; their result is consistent with our
  prediction.  These results may indicate that our model does not
  underestimate dust attenuation of galaxies at $z\gtrsim 1$ but
  overestimates it at $z\approx 0$.  This interpretation will be
  examined as in our future work.

  \subsection{Comparison with the HB06 CSFRD}
  \label{subsec:Comments2HB06}

  As described in Section~\ref{subsec:CorrDustAtten}, HB06 obtained
  the CSFRD at $z < 3$ by summing the UV data and FIR measurements.
  They reported that this technique of UV$+$FIR measurements gives an
  effective obscuration correction to the UV data by a factor of 2 at
  $z\approx 0$ and $\sim 5$ (i.e., 1.7 mag) at $z \gtrsim 1$.  It may
  be difficult to reconcile this with our result and should be
  examined in detail.

  However, the correction factor of $\sim 5$ at $z\gtrsim 1$ reported
  in HB06 might be overestimated for the following reason.  HB06
  performed the effective obscuration correction by just adding a
  constant CSFRD measured by Le Floc'h et al. (2005) for FIR
  wavelengths at $z=1$.  This is based on the observational result
  that FIR measurements of the CSFRD are quite flat in the range
  $z=1$--3 as reported by P\'{e}rez-Gonz\'{a}lez et al. (2005), who
  estimated the total IR luminosity of each galaxy by using the local
  template spectral energy distribution of Chary \& Elbaz (2001).
  However, as reported by Murphy et al. (2011), using this template
  results in an overestimation of the total IR luminosity for
  IR-bright galaxies at $z>1.5$.  Hence, the IR measurement of
  P\'{e}rez-Gonz\'{a}lez et al. (2005) may be overestimated.  We are
  planning to investigate whether our model reproduces the observed IR
  data in future studies.

  \subsection{Empirical Calibrations of the Obscuration Correction and
  the SFR Conversion from UV Luminosity}
  \label{subsec:empirical-calib}

   Here we propose new empirical formulas which correct for dust
   obscuration and convert from the intrinsic UV LD,
   $\rho_\mathrm{UV}^\mathrm{int}$, to the CSFRD, $\dot{\rho}_\star$.
   These formulas are derived to reproduce the true quantities for the
   model galaxies and are represented by explicit analytic functions.
   For the obscuration correction, we show two different formulas; one
   is for the observable LF and the other is for its integrated
   quantity $\rho_\mathrm{UV}$.  These two formulas are, respectively,
   similar to the SFR-dependent and common corrections of H04, but
   they are described to have redshift dependence.

   \subsubsection{Conversion from Observable LF into Intrinsic LF}

\begin{figure}
    \epsscale{1.15}
    \plotone{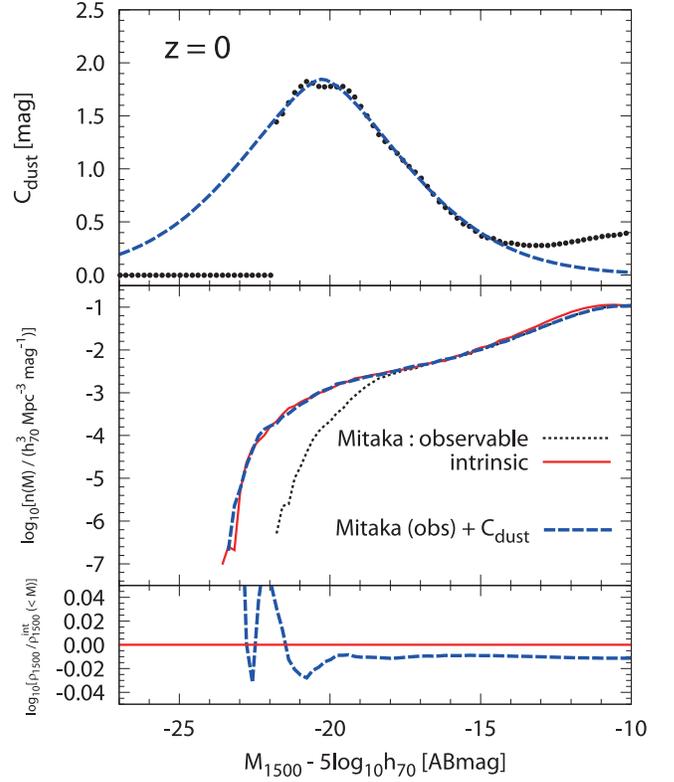}
    \caption{
    Top, middle, and bottom panels show $C_\mathrm{dust}$, LF, and the
    ratio of the LD calculated from the Mitaka observable LF
    $+C_\mathrm{dust}$ to the Mitaka intrinsic LD, respectively, at
    $z=0$ for a wavelength of $\lambda = 1500$~{\AA}.  In the top
    panel, the filled circles are the numerical data calculated from
    the intrinsic and observable 1500~{\AA} LFs of the Mitaka model,
    while the dashed curve is analytical fit using
    Equation~(\ref{eq-Cdust-fit}) with the numerical quantities in
    Table~\ref{tab-Cdust}.  $C_\mathrm{dust} = 0$ at $M_{1500}
    \lesssim -22$~mag is reflected by the fact that there is no model
    galaxy with such bright observable magnitudes.  In the middle
    panel, the solid and dotted curves are the intrinsic and
    observable LFs of the Mitaka model, respectively, while the dashed
    curve is the Mitaka observable LF + $C_\mathrm{dust}$.
    }
    \label{fig-Cdust-z0}
   \end{figure}

   Let us define $C_\mathrm{dust}$ as an empirical formula to convert
   the observable LF into an intrinsic one.  $C_\mathrm{dust}$ at a
   certain magnitude $M_{1500}$ is determined via an
   abundance-matching approach.  That is, the cumulative number
   density of the observable LF of the Mitaka model, $n^\mathrm{obs}(<
   M_{1500})$, should match that of the intrinsic LF,
   $n^\mathrm{int}(< M_{1500} - C_\mathrm{dust})$:
   \begin{equation}
    n^\mathrm{obs}(< M_{1500}) = n^\mathrm{int}(<
     M_{1500} - C_\mathrm{dust}).
   \end{equation}
   The top panel of Figure~\ref{fig-Cdust-z0} shows the resultant
   $C_\mathrm{dust}$ as a function of $M_{1500}$ at $z=0$.

   We have derived $C_\mathrm{dust}$ for other redshifts and found
   that $C_\mathrm{dust}$ can be fitted well with the following
   analytic function in the redshift range $z=0$--10:
   \begin{equation}
    C_\mathrm{dust}(M_{1500};\ z)
     = a\ \exp{\left[-b |M_{1500} - M_{1500}^0|^c\right]}.
     \label{eq-Cdust-fit}
   \end{equation}
   Here $a,\ b,\ c,$ and $M_{1500}^0$ are model parameters and evolve
   with redshift.  $a$ and $M_{1500}^0$ have units of magnitude, while
   $b$ and $c$ are non-dimensional constants.  We adopt a smoothly
   declining functional form even for the bright magnitudes where
   there are no model galaxies.  We note that $C_\mathrm{dust}$ is a
   dust obscuration correction for the UV LF as a statistical quantity
   like the SFR-dependent correction of H04.  Hence it does not
   represent a mean dust attenuation for the galaxies at a certain
   magnitude.  Actually, the intrinsically UV-brightest galaxies are
   not the observationally brightest galaxies in our model, as
   described in Section 4.1.  The reason why $C_\mathrm{dust}$ has a
   peak around the characteristic magnitude $M_{1500}^\ast$ is that
   the magnitude difference between the intrinsic and observable UV
   LFs becomes largest at $M_{1500} \approx M_{1500}^\ast$, not that
   the mean attenuation of the galaxies at $M_{1500} \approx
   M_{1500}^\ast$ is the largest.

   There are significant discrepancies between the raw
   $C_\mathrm{dust}$ quantities and that evaluated using
   Equation~(\ref{eq-Cdust-fit}), as represented by the filled circles
   and dashed curve, respectively, in the top panel of
   Figure~\ref{fig-Cdust-z0} at $M_{1500} \lesssim -23$~mag or
   $M_{1500} \gtrsim -15$~mag.  Fortunately, this deviation hardly
   affects the corrected UV LF or the integrated LD as shown in the
   middle and bottom panels of Figure~\ref{fig-Cdust-z0}.
   Table~\ref{tab-Cdust} gives the numerical quantities for the
   best-fit parameters in the redshift range $z = 0$--10.
   \begin{deluxetable}{ccccc}
    \tablewidth{250pt}
    \tabletypesize{\scriptsize}
    \tablecolumns{5}
    \tablecaption{Fitting Parameters for our Formula to Correct Dust
    Attenuation, $C_\mathrm{dust}$}
    \tablehead{
    Redshift, $z$ & $a$ & $b$ & $c$ & $M_{1500}^0$\\
    & (mag) & & & (mag)
    }
    \startdata
    0  & $1.847$ & $0.1139$ & $1.566$ & $-19.49$ \\
    1  & $2.466$ & $0.1181$ & $1.281$ & $-22.65$ \\
    2  & $2.344$ & $0.2625$ & $1.086$ & $-21.86$ \\
    3  & $2.235$ & $0.2844$ & $1.184$ & $-21.91$ \\
    4  & $2.032$ & $0.2115$ & $1.428$ & $-22.15$ \\
    5  & $1.822$ & $0.1971$ & $1.599$ & $-22.16$ \\
    6  & $1.583$ & $0.1673$ & $1.737$ & $-22.14$ \\
    7  & $1.496$ & $0.1630$ & $1.800$ & $-22.08$ \\
    8  & $1.415$ & $0.1900$ & $1.641$ & $-22.16$ \\
    9  & $1.228$ & $0.1239$ & $1.895$ & $-22.39$ \\
    10 & $1.109$ & $0.1898$ & $1.634$ & $-22.17$
    \enddata
    \tablecomments{The analytic expression for $C_\mathrm{dust}$ is
    given by Equation~(\ref{eq-Cdust-fit}).}
    \label{tab-Cdust}
   \end{deluxetable}

   \begin{deluxetable}{cccccc}
    \tabletypesize{\scriptsize}
    \tablewidth{\linewidth}
    \tablecolumns{6}
    \tablecaption{Fitting Parameters for our Formulas for
    $C_\mathrm{dust}^\mathrm{eff}$ and $C_\mathrm{SFR}^\mathrm{eff}$\label{tab-FitParameters}}
    \tablehead{
    \multicolumn{2}{c}{\scriptsize $C_\mathrm{dust}^\mathrm{eff}$} & \colhead{} &
    \multicolumn{3}{c}{\scriptsize $C_\mathrm{SFR}^\mathrm{eff}$}\\
    \cline{1-2} \cline{4-6}\\
    {\scriptsize $\alpha$} & {\scriptsize $\beta$} & \colhead{} & {\scriptsize $C_2$} & {\scriptsize $C_1$} & {\scriptsize $C_0$} \\
    \colhead{} & \colhead{} & \colhead{} & \colhead{} & \colhead{} &
    {\scriptsize ($M_\odot \ \mathrm{yr^{-1}\ (ergs/ s/ Hz)^{-1}}$)}
    }
    \startdata
    {\scriptsize $2.983$} & {\scriptsize $0.3056$} & & {\scriptsize $-5.915\times 10^{-5}$} & {\scriptsize $7.294\times 10^{-4}$} & {\scriptsize $-28.01$}
    \enddata
    \tablecomments{The analytic expressions for
    $C_\mathrm{dust}^\mathrm{eff}$ and $C_\mathrm{SFR}^\mathrm{eff}$ are given by
    Equations~(\ref{eq-CdustEff}) and (\ref{eq-Lnu2SFR}), respectively.}
   \end{deluxetable}

   The normalization factor $a$ indicates the maximum value of
   $C_\mathrm{dust}$ at the redshift.  $a$ is found to gradually
   increase with redshift toward its peak at $z\simeq 1$--3 and then
   decreases as $z$ increases.  It is interesting that the peak
   redshift for $a$ is roughly equal to the redshift where dusty
   galaxies (e.g., ultraluminous infrared galaxies, sub-mm galaxies,
   etc.) are mainly found.

   \subsubsection{Conversion from Observable UV LD into
   Intrinsic UV LD}

   The conversion factor from the observed 1500~{\AA} LD,
   $\rho_{1500}$, into the intrinsic one, $\rho_{1500}^\mathrm{int}$,
   is defined as $C_\mathrm{dust}^\mathrm{eff} \equiv
   \rho_{1500}^\mathrm{int} / \rho_{1500}$.  This relates to the
   effective dust attenuation $A_{1500}^\mathrm{eff}$ defined in
   Equation~(\ref{eq-Alameff}) via $C_\mathrm{dust}^\mathrm{eff} =
   \mathrm{dex}( 0.4 A_{1500}^\mathrm{eff}
   )$.\footnote{$\mathrm{dex}(x)$ is the inverse function of
   $\log_{10}(x)$: $\mathrm{dex}(x) \equiv 10^x$.}  It is found that
   the $C_\mathrm{dust}^\mathrm{eff}$ for our model galaxies can be
   well fitted by the following simple analytic function with two
   redshift-independent parameters $\alpha$ and $\beta$ in the
   redshift range $z = 0$--10 as shown in Figure~\ref{fig-Alameff-z}:
   \begin{equation}
    C_\mathrm{dust}^\mathrm{eff}(z) = \alpha \ \exp{
     \left[-\beta \left(1+z \right)\right]} + 1.
     \label{eq-CdustEff}
   \end{equation}
   This functional form is motivated by the natural expectation that
   $C_\mathrm{dust}^\mathrm{eff}$ approaches unity at high redshift.
   Since there is little dust at high redshift,
   $C_\mathrm{dust}^\mathrm{eff} \approx 1$.  The best-fit parameters
   which reproduce the ratio $\rho_{1500}^\mathrm{int} / \rho_{1500}$
   at $z=0$--10 within $\pm 10\%$ are given in
   Table~\ref{tab-FitParameters}.

   While the normalization parameter $a$ in
   Equation~(\ref{eq-Cdust-fit}) has its peak at $z\simeq 1$--3,
   $C_\mathrm{dust}^\mathrm{eff}$ decreases monotonically with
   increasing $z$.  The different redshift evolution can be
   interpreted by the fact that the contribution from galaxies fainter
   than $M_{1500}^0$, where the galaxies have a maximum extinction of
   $a$ at the redshift, to the intrinsic 1500~{\AA} LD is significant;
   for a typical faint-end slope $\alpha \approx -1.4$ of the UV LF,
   the contribution from galaxies with $L \lesssim L_\ast $ reaches
   $\sim 0.7$~dex as shown in the right panel of
   Figure~\ref{fig-Schechter}.  Since such faint galaxies have smaller
   $C_\mathrm{dust}$ than its peak value of $a$,
   $C_\mathrm{dust}^\mathrm{eff}$ progressively decreases toward high
   redshift although $a$ has its peak at $z\simeq 1$--3.

   \subsubsection{Conversion from Intrinsic UV LD into CSFRD}

   We find that the ratio of the CSFRD to the intrinsic 1500~{\AA} LD,
   $\dot{\rho}_\star / \rho_{1500}^\mathrm{int}$, can be well fitted
   with the following simple quadratic function in the redshift range
   $z=0$--10:
   \begin{equation}
    C_\mathrm{SFR}^\mathrm{eff}(z)
     = C_0\left[1 + C_1\left(1+z\right) + C_2\left(1+z\right)^2\right].
     \label{eq-Lnu2SFR}
   \end{equation}
   Here $C_0,\ C_1,$ and $C_2$ are the model parameters and are
   redshift-independent constants.  $C_0$ has the same dimensions as
   $C_\mathrm{SFR}^\mathrm{eff}$, $M_\odot\ \mathrm{yr^{-1}\ (erg\
   s^{-1}\ Hz^{-1})^{-1}}$, while $C_1$ and $C_2$ are non-dimensional
   constants.  With the best-fit quantities given in
   Table~\ref{tab-FitParameters}, the analytic function reproduces our
   model results for $\dot{\rho}_\star / \rho_{1500}^\mathrm{int}$
   within $\pm 5$\%.

   We note here that, while the empirical formula given in
   Equation~(\ref{eq-Lnu2SFR}) predicts that
   $C_\mathrm{SFR}^\mathrm{eff}$ has a peak at $z\sim 6$ and
   progressively decreases toward high redshifts, it is simply a
   fitting result and does not have any physical motivation.
   Nevertheless, it might be a real trend as discussed in
   Section~\ref{subsec-ResCSFR}.

  \subsection{Recipe for Converting Observed UV LF into CSFRD}

  Here we describe a recipe for converting the observed 1500~{\AA} LF
  into the CSFRD $\dot{\rho}_\star$ using our empirical formulas given
  in Section~\ref{subsec:empirical-calib}.  As a first step, the
  intrinsic 1500~{\AA} LD, $\rho_{1500}^\mathrm{int}$, should be
  calculated from the observed UV LF.  This can be done using either
  one of the following two approaches.  The first approach is to
  convert the observed UV LF into an intrinsic LF via the empirical
  formula for $C_\mathrm{dust}$ as a function of magnitude and
  redshift given in Equation~(\ref{eq-Cdust-fit}) with the best-fit
  parameters in Table~\ref{tab-Cdust}.  Then the intrinsic UV LF is
  integrated over magnitude to obtain the intrinsic UV LD,
  $\rho_{1500}^\mathrm{int}$.  The second approach is to first
  integrate the observed UV LF over magnitude to obtain the observable
  UV LD, $\rho_{1500}$, and then converted it into
  $\rho_{1500}^\mathrm{int}$ using the empirical formula for
  $C_\mathrm{dust}^\mathrm{eff}$ as a function of redshift given in
  Equation~(\ref{eq-CdustEff}).  Finally, one can evaluate
  $\dot{\rho}_\star$ from $\rho_{1500}^\mathrm{int}$ using the
  empirical formula for $C_\mathrm{SFR}^\mathrm{eff}$ as a function of
  redshift given in Equation~(\ref{eq-Lnu2SFR}).

  For $C_\mathrm{dust}$ given in Equation~(\ref{eq-Cdust-fit}), it is
  statistically enough to linearly interpolate the formula for a
  specified redshift while the best-fit parameters are provided
  discretely in redshift for our empirical formula.  One should
  interpolate the parameters to evaluate an adequate $C_\mathrm{dust}$
  at a certain magnitude and the desired redshift rather than
  interpolating $C_\mathrm{dust}$ itself.

  \begin{deluxetable*}{ccccc}
   \tabletypesize{\scriptsize}
   \tablewidth{0pt}
   \tablecolumns{5}
   \tablecaption{LD and CSFRD of the Mitaka Model}
   \tablehead{
   Redshift, $z$ & $\dot{\rho_\star}$ & $\rho_{1500}$ & $\rho_{1500}^\mathrm{int}$ & $\dot{\rho_\star} / \rho_{1500}$\\
   & $(h_{70}\ M_\odot\ \mathrm{yr^{-1}\ Mpc^{-3}})$ & $(h_{70}\ \mathrm{erg\ s^{-1}\ Hz^{-1}\ Mpc^{-3}})$ & $(h_{70}\ \mathrm{erg\ s^{-1}\ Hz^{-1}\ Mpc^{-3}})$ & $(M_\odot\ \mathrm{yr^{-1}\ (erg\ s^{-1}\ Hz^{-1})^{-1}})$}
   \startdata
   0 & $2.599\times 10^{-2}$ & $8.090\times 10^{25}$ & $2.616\times 10^{26}$ & $3.213\times 10^{-28}$ \\
   1 & $6.825\times 10^{-2}$ & $2.531\times 10^{26}$ & $6.593\times 10^{26}$ & $2.697\times 10^{-28}$ \\
   2 & $8.285\times 10^{-2}$ & $3.665\times 10^{26}$ & $7.910\times 10^{26}$ & $2.261\times 10^{-28}$ \\
   3 & $6.828\times 10^{-2}$ & $3.498\times 10^{26}$ & $6.396\times 10^{26}$ & $1.952\times 10^{-28}$ \\
   4 & $4.926\times 10^{-2}$ & $2.752\times 10^{26}$ & $4.518\times 10^{26}$ & $1.790\times 10^{-28}$ \\
   5 & $3.124\times 10^{-2}$ & $1.902\times 10^{26}$ & $2.818\times 10^{26}$ & $1.642\times 10^{-28}$ \\
   6 & $1.759\times 10^{-2}$ & $1.119\times 10^{26}$ & $1.540\times 10^{26}$ & $1.572\times 10^{-28}$ \\
   7 & $9.105\times 10^{-3}$ & $5.988\times 10^{25}$ & $7.849\times 10^{25}$ & $1.521\times 10^{-28}$ \\
   8 & $4.524\times 10^{-3}$ & $3.129\times 10^{25}$ & $3.865\times 10^{25}$ & $1.446\times 10^{-28}$ \\
   9 & $2.069\times 10^{-3}$ & $1.641\times 10^{25}$ & $1.898\times 10^{25}$ & $1.261\times 10^{-28}$ \\
   10& $9.514\times 10^{-4}$ & $8.245\times 10^{24}$ & $9.150\times 10^{24}$ & $1.154\times 10^{-28}$
   \enddata
   \tablecomments{$\dot{\rho}_\star$ is the CSFRD and
   $\rho_{1500}$ and $\rho_{1500}^\mathrm{int}$ represent the observable
   and intrinsic 1500~{\AA} LDs, respectively, for all of the galaxies
   in the Mitaka model.}
   \label{tab-CSFRDdata}
  \end{deluxetable*}

  \begin{deluxetable*}{lccr}
   \tabletypesize{\scriptsize}
   \tablecolumns{4}
   \tablecaption{CSFRD Parametric Fits to Various Forms from the Literatures}
   \tablehead{
   Reference & Functional Form & Parameter & Value
   }
   \startdata
   Cole et al. (2001) & $\dot{\rho}_\ast(z) =
   \left(a+bz\right)h\ /\ [1+\left(z/c\right)^d]$ & $a$ & $0.0389$ \\
   & & $b$ & $0.0545$ \\
   & & $c$ & $2.973$ \\
   & & $d$ & $3.655$ \\
   \hline
   Hernquist \& Springel (2003) & $\dot{\rho}_\ast(z)
   = \dot{\rho}_0\chi^2\ /\ [1+\alpha (\chi-1)^3\exp{(\beta
   \chi^{7/4})}]$ & $\dot{\rho}_0$ & $0.030$\\
   & $\chi=[H(z)/H_0]^{2/3}$ & $\alpha$ & $0.323$\\
   & & $\beta$ & $0.051$\\
   \hline
   Y\"{u}ksel et al. (2008) & $\dot{\rho}_\ast(z) = \dot{\rho}_0
   [(1+z)^{\alpha \eta}+\{(1+z)/B\}^{\beta \eta}+\{(1+z)/C\}^{\gamma
   \eta}]^{1/\eta}$, & $\dot{\rho}_0$ &  $0.0258$ \\
   & $B=(1+z_1)^{1-\alpha /\beta}$,
   \phm{...}$C=(1+z_1)^{(\beta -\alpha)/\gamma}(1+z_2)^{1-\beta/\gamma}$
   & $\alpha$   &  $1.6$ \\
   & & $\beta$  & $-1.2$ \\
   & & $\gamma$ & $-5.7$ \\
   & & $z_1$ &  $1.7$ \\
   & & $z_2$ &  $5.0$ \\
   & & $\eta$ & $-1.62$
   \enddata
   \label{tab-CSFRDfits}
  \end{deluxetable*}

  The numerical quantities for $\dot{\rho}_\star,\ \rho_{1500},$ and
  $\rho_{1500}^\mathrm{int}$ of the Mitaka model are compiled in
  Table~\ref{tab-CSFRDdata}.  We also present the CSFRD parametric
  fits to a variety of analytic forms in the literature (Cole et
  al. 2001; Hernquist \& Springel 2003; Y\"{u}ksel et al. 2008) in
  Table~\ref{tab-CSFRDfits}.
\acknowledgments

We thank Shunsaku Horiuchi and John Beacom for useful discussions.  We
thank the referee for his/her many helpful comments and suggestions
which improved this paper.  The numerical calculations were in part
carried out on the general-purpose PC farm at Center for Computational
Astrophysics, CfCA, of National Astronomical Observatory of Japan.
M.A.R.K. and YI were supported by the Research Fellowship for Young
Scientists from the Japan Society for the Promotion of Science (JSPS).
A.K.I. is supported by JSPS KAKENHI 23684010.


\begin{thebibliography}{}

\bibitem[Benson(2012)]{b12}
Benson, A. J.
2012, NewA, 17, 175

\bibitem[Bond et al.(1991)]{bond91}
Bond, J. R., Cole, S., Efstathiou, G., \& Kaiser, N.
1991, \apj, 379, 440

\bibitem[Botticella et al.(2012)]{b12}
Botticella, M. T., Smartt, S. J., Kennicutt, R. C., Jr., et al.
2012, \aap, 537, A132

\bibitem[Bower(1991)]{bower91}
Bower, R.
1991, \mnras, 248, 332

\bibitem[Buat et al.(2012)]{buat12}
Buat, V., Noll, S., Burgarella, D., et al.
2012, \aap, 545, A141

\bibitem[Burgarella et al.(2005)]{burgarella05}
Burgarella, D., Buat, V., \& Iglesias-P\'{a}ramo
2005, \mnras, 360, 1413

\bibitem[Calzetti et al.(2000)]{calzetti00}
Calzetti, D., Armus, L., Bohlin, R. C., et al.
2000, \apj, 533, 682

\bibitem[Calzetti et al.(1994)]{calzetti94}
Calzetti, D., Kinney, A. L., \& Storchi-Bergmann, T.
1994, \apj, 429, 582

\bibitem[Casey et al.(2012)]{casey12}
Casey, C. M., Berta, S., B\'{e}thermin, M., et al.
2012, \apj, 761, 140

\bibitem[Chary \& Elbaz(2001)]{ce01}
Chary, R. \& Elbaz, Z.
2001, \apj, 556, 562

\bibitem[Choi \& Nagamine(2011)]{cn11}
Choi, J.-H., \& Nagamine, K.
2012, \mnras, 419, 1280

\bibitem[Clemens \& Alexander(2004)]{ca04}
Clemens, M. S. \& Alexander, P.
2004, \mnras, 350, 66

\bibitem[Cole et al.(2001)]{cole01}
Cole, S., Norberg, P., Baugh, C. M., et al.
2001, \mnras, 326, 255

\bibitem[Coward et al.(2008)]{co08}
Coward, D. M., Guetta, D., Burman, R. R., \& Imerito, A.
2008, \mnras, 386, 111

\bibitem[Cowie et al.(1996)]{cowie96}
Cowie, L. L., Songaila, A., Hu, E. M., \& Cohen, J. G.
1996, \aj, 112, 839

\bibitem[Cucciati et al.(2012)]{c12}
Cucciati, O., Tresse, L., Ilbert, O., et al.
2012, \aap, 539, 31

\bibitem[Fioc \& Rocca-Volmerange(1997)]{fr97}
Fioc, M., \& Rocca-Volmerange, B.
1997, \aap, 326, 950

\bibitem[Garnett(2002)]{garnett02}
Garnett, D. R.
2002, \apj, 581, 1019

\bibitem[Hernquist \& Springel(2003)]{2003MNRAS.341.1253H}
Hernquist, L., \& Springel, V.
2003, \mnras, 341, 1253

\bibitem[Hopkins (2004)]{hop04}
Hopkins, A. M.
2004, \apj, 615, 209 (H04)

\bibitem[Hopkins \& Beacom(2006)]{hb06}
Hopkins, A. M., \& Beacom, J. F.
2006, \apj, 651, 142 (HB06)

\bibitem[Hopkins et al.(2001)]{hop01}
Hopkins, A. M., Connolly, A. J., Haarsma, D. B., \& Cram, L. E.
2001, \aj, 122, 288

\bibitem[Horiuchi et al.(2011)]{horiuchi11}
Horiuchi, S., Beacom, J. F., Kochanek, C. S., et al.
2011, \apj, 738, 154

\bibitem[Karim et al.(2011)]{karim11}
Karim, A., Schinnerer, E., Mart\'{i}nez-Sansigre, A., et al.
2011, \apj, 730, 61

\bibitem[Kashikawa et al.(2006)]{2006ApJ...637..631K}
Kashikawa, N., Yoshida, M., Shimasaku, K., et al.\ 2006, \apj, 637, 631

\bibitem[Kennicutt(1998)]{k98}
Kennicutt, R. C., Jr.
1998, \araa, 36, 189 (K98)


\bibitem[Kistler et al.(2009)]{kistler09}
Kistler, M. D., Y\"{u}ksel, H., Beacom, J. F., Hopkins, A. M., \&
		Wyithe, J. S. B.
2009, \apj, 705, 104

\bibitem[Kobayashi et al.(2007)]{mark07}
Kobayashi, A. R. M., Totani, T., \& Nagashima, M.
2007, \apj, 670, 919

\bibitem[Kobayashi et al.(2010)]{mark10}
Kobayashi, A. R. M., Totani, T., \& Nagashima, M.
2010, \apj, 708, 1119

\bibitem[Kodama \& Arimoto(1997)]{ka97}
Kodama, T. \& Arimoto, N.
1997, \aap, 320, 41

\bibitem[Lacey et al.(2010)]{l10}
Lacey, C. G., Baugh, C. M., Frenk, C. M., \& Benson, A. J.
2011, \mnras, 412, 1828

\bibitem[Lacey \& Cole(1993)]{lc93}
Lacey, C. G., \& Cole, S.
1993, \mnras, 262, 627

\bibitem[Le Floc'h et al.(2005)]{lf05}
Le Floc'h, E., Papovich, C., Dole, H., et al.
2005, \apj, 632, 169

\bibitem[Lilly et al.(1996)]{lilly96}
Lilly, S. J., Le F\'{e}vre, O., Hammer, F., \& Crampton, D.
1996, ApJL, 460, 1

\bibitem[Madau et al.(1996)]{madau96}
Madau, P., Ferguson, H. C., Dickinson, M. E., et al.
1996, \mnras, 283, 1388

\bibitem[Madau et al.(1998)]{madau98}
Madau, P., Pozzetti, L., \& Dickinson, M.
1998, \apj, 498, 106

\bibitem[Murphy et al.(2011)]{murphy11}
Murphy, E. J., Chary, R.-R., Dickinson, M., et al.
2011, \apj, 732, 126

\bibitem[Nagamine et al.(2006)]{2006ApJ...653..881N}
Nagamine, K., Ostriker, J.~P., Fukugita, M., \& Cen, R.\ 2006, \apj, 653, 881

\bibitem[Nagashima et al.(2005)]{nagashima05b}
Nagashima, M., Yahagi, H., Enoki, M., Yoshii, Y., \& Gouda, N.
2005, \apj, 634, 26

\bibitem[Nagashima \& Yoshii(2004)]{ny04}
Nagashima, M., \& Yoshii, Y.
2004, \apj, 610, 23

\bibitem[Oesch et al.(2010)]{oesch10}
Oesch, P. A., Bouwens, R. J., Carollo, C. M., et al.
2010, ApJL, 725, 150

\bibitem[Pei(1992)]{pei92}
Pei, Y. C.
1992, \apj, 395, 130

\bibitem[P\'{e}rez-Gonz\'{a}lez et al.(2005)]{pg05}
P\'{e}rez-Gonz\'{a}lez, P. G., Rieke, G. H., Egami, E., et al.
2005, \apj, 630, 82

\bibitem[Raue \& Meyer(2012)]{rm12}
Raue, M. \& Meyer, M.
2012, \mnras, 426, 1097

\bibitem[Reddy \& Steidel(2009)]{rs09}
Reddy, N. A., \& Steidel, C. C.
2009, \apj, 692, 778

\bibitem[Rela\~{n}o et al.(2012)]{relano12}
Rela\~{n}o, M., Kennicutt, R. C., Jr., Eldridge, J. J., Lee, J. C., \&
		Verley, S.
2012, \mnras, 423, 2933

\bibitem[Robotham \& Driver(2011)]{rd11}
Robotham, A. S. G., \& Driver, S. P.
2011, \mnras, 413, 2570

\bibitem[Schaerer(2003)]{schaerer03}
Schaerer, D.
2003, \aap, 397, 527

\bibitem[Strigari et al.(2005)]{strigari05}
Strigari, L. E., Beacom, J. F., Walker, T. P., \& Zhang, P.
2005, J. Cosmol. Astropart. Phys., JCAP04(2005)017

\bibitem[Takeuchi et al.(2005)]{takeuchi05}
Takeuchi, T. T., Buat, V., \& Burgarella, D.
2005, \aap, 440, L17

\bibitem[Tominaga et al.(2011)]{tominaga11}
Tominaga, N., Morokuma, T., Blinnikov, S. I., et al.
2011, \apjs, 193, 20

\bibitem[Wang \& Dai(2011)]{wd11}
Wang, F. Y., \& Dai, Z. G.
2011, \apj, 727, 34

\bibitem[Wilkins et al.(2008)]{wilkins08}
Wilkins, S. M., Trentham, N., \& Hopkins, A. M.
2008, \mnras, 385, 687

\bibitem[Wyder et al.(2005)]{wyder05}
Wyder, T. K., Treyer, M. A., Milliard, B., et al.
2005, ApJL, 619, 15

\bibitem[Yabe et al.(2009)]{2009ApJ...693..507Y}
Yabe, K., Ohta, K., Iwata, I., et al.\ 2009, \apj, 693, 507

\bibitem[Yahagi et al.(2004)]{yny04}
Yahagi, H., Nagashima, M., \& Yoshii, Y.
2004, \apj, 605, 709

\bibitem[Y\"{u}ksel et al.(2008)]{y08}
Y\"{u}ksel, H., Kistler, M. D., Beacom, J. F., \& Hopkins, A. M.
2008, ApJL, 683, 5

\end{thebibliography}
\end{document}